\documentclass[iop]{emulateapj}
\slugcomment{Accepted by The Astrophysical Journal}

\usepackage{fancyhdr}
\usepackage{amsmath}
\DeclareMathOperator{\sinc}{sinc}
\DeclareMathOperator{\re}{Re}
\DeclareMathOperator{\im}{Im}
\usepackage[mathscr]{euscript}

\shorttitle{Systematic effects in polarizing FTSs for CMB observations}
\shortauthors{Nagler \sl{et. al}}

\begin{document}

\title{Systematic effects in polarizing Fourier transform spectrometers \\ for cosmic microwave background observations}


\author{Peter C. Nagler}
\affil{Department of Physics, Brown University, Providence, RI 02912, USA}
\affil{NASA/Goddard Space Flight Center, Code 553, Greenbelt, MD 20771, USA}
\email{Email: peter.c.nagler@nasa.gov}

\author{Dale J. Fixsen}
\affil{NASA/Goddard Space Flight Center, Code 665, Greenbelt, MD 20771, USA}

\author{Alan Kogut}
\affil{NASA/Goddard Space Flight Center, Code 665, Greenbelt, MD 20771, USA}

\and

\author{Gregory S. Tucker}
\affil{Department of Physics, Brown University, Providence, RI 02912 USA}

\begin{abstract}
The detection of the primordial {\sl B}-mode polarization signal of the cosmic microwave background (CMB) would provide evidence for inflation. Yet as has become increasingly clear, the detection of a such a faint signal requires an instrument with both wide frequency coverage to reject foregrounds and excellent control over instrumental systematic effects. Using a polarizing Fourier transform spectrometer (FTS) for CMB observations meets both these requirements. In this work, we present an analysis of instrumental systematic effects in polarizing Fourier transform spectrometers, using the Primordial Inflation Explorer (PIXIE) as a worked example. We analytically solve for the most important systematic effects inherent to the FTS - emissive optical components, misaligned optical components, sampling and phase errors, and spin synchronous effects - and demonstrate that residual systematic error terms after corrections will all be at the sub-nK level, well below the predicted 100 nK {\sl B}-mode signal.
\end{abstract}


\section{Introduction}

Fourier transform spectrometer (FTS) measurements from balloon \citep{Woodyetal1981, Tuckeretal1997}, sounding rocket \citep{Gushetal1990}, and satellite \citep{Fixsenetal1996} platforms have made significant contributions to cosmic microwave background (CMB) studies. Recently there has been renewed interest in deploying a FTS from a satellite platform \citep[e.g.][]{Kogutetal2011, Kogutetal2014, Andreetal2014} in order to measure the {\sl B}-mode polarization signal of the CMB \citep{Kamionkowskietal1997, Seljaketal1997}. Such a measurement would provide a critical test of inflationary cosmology, probing physics at energies twelve orders of magnitude higher than is accessible to particle accelerators.

FTSs are particularly well equipped to measure the faint {\sl B}-mode signal. First, they cover a wide frequency band with many channels, enabling the measurement and rejection of foreground signals. This is critical in order to distinguish between cosmological signals and signals originating within our own galaxy \citep[e.g.][]{BICEP2015}. Second, their symmetry leads to excellent control over instrumental systematic effects, the subject of this paper. Using the Primordial Inflation Explorer (PIXIE) as a worked example \citep{Kogutetal2011, Kogutetal2014}, we demonstrate that instrumental systematic effects from a FTS can be mitigated to the sub-nK level, well below the predicted 100 nK {\sl B}-mode signal.

The PIXIE experiment is based on a simple ansatz. The CMB is a near-perfect blackbody at a temperature of 2.725 K. The PIXIE instrument will be kept isothermal with the CMB at a temperature of 2.725 K. Therefore the detectors will always be looking into a near-perfect blackbody cavity at the CMB temperature. Regardless of where the instrument points or of internal absorption and emission, the detectors will see the same thing, a Planck spectrum with a temperature of 2.725 K. Photons incident from the sky will be indistinguishable from those emitted by the instrument.

PIXIE is a space-based instrument with two input beams that are co-aligned and the entire instrument spins about the same axis with a period of $\sim$ 15 seconds. Thus in a few seconds the polarization directions are interchanged. The instrument spin axis moves so that over the course of a few hours a great circle is traced out on the sky. This circle, 90 degrees from the sun, precesses so that over the course of a year the entire sky is mapped
twice.

As a nulling interferometer, PIXIE is sensitive only to the difference between orthogonal polarizations of incident light. Therefore to the extent the CMB is a blackbody, and to the extent its optics are isothermal with the CMB, PIXIE will measure zero. As a result, the PIXIE design is ideal for measuring faint polarized signals in a bright unpolarized background.

The same concept applies to systematic error sourced from imperfections in the PIXIE FTS. In general, the instrument is not sensitive to absolute non-idealities, but instead differential non-idealities between symmetrically positioned optical components. Descriptions of the ideal instrument are available \citep{Kogutetal2011, Kogutetal2014}, but we review it and introduce analytic methods in Section \ref{sec:Ideal Instrument}.

The systematic effects are separated into several categories. The first, denoted emission errors, result from emissive optical components that absorb and emit radiation. These are treated in Section \ref{sec:electromagnetic-error}. The second, denoted geometric errors, come about when a given optical component is not perfectly aligned, generally leading to reductions in optical efficiency. These are treated in Section \ref{sub:Geometric-errors}. The third, denoted mirror transport mechanism (MTM) errors, result from systematic offsets and uncertainties in the moving mirror assembly's position. These are treated in Section \ref{sec:MTMerr}. Finally in Section \ref{sec:Spin-synchronous-effects} we treat spacecraft spin-synchronous effects, which can cause various instrumental drifts. For systematic errors of each kind, we solve for their analytic form, estimate their uncorrected magnitude, and subsequently show how they can be corrected to below $1$ nK. 

The list of systematic errors covered in this work is not exhaustive but it includes the most important effects unique to the FTS nature of PIXIE and other similar instruments. Analyses of other effects that are common to many instrument types, such as beam effects on the sky, foreground subtraction, and particle hits on the detectors, will be treated in separate works. We explicitly exclude beam effects from this paper. These are largely determined by the fore optics which are better treated in an optical paper.

\section{The ideal instrument}\label{sec:Ideal Instrument}

\subsection{Optical design}

\begin{figure}[t]
\centering
\includegraphics[width=1\columnwidth]{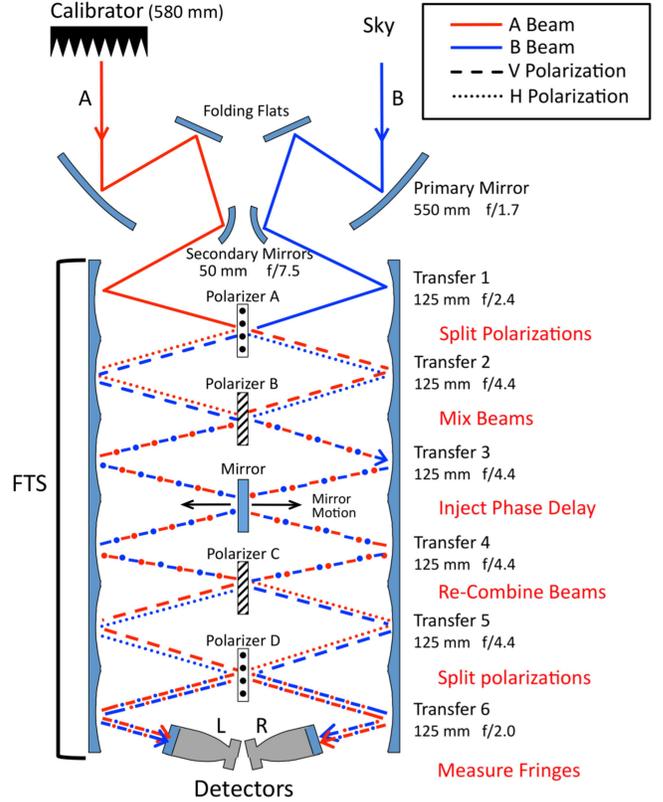}
\caption{Cartoon of the PIXIE FTS. The listed dimensions give the diameters of components. The calibrator can block either beam or be stowed. The beams are co-pointed. Polarizers A and D are oriented with their grids vertical, transmitting horizontal ($\hat{x}$) polarization and reflecting vertical ($\hat{y}$) polarization. Polarizers B and C are oriented at 45 degrees relative to polarizers A and D. The mirror stroke is $\pm2.6$ mm, which corresponds to an optical path difference between beams of $\pm10.0$ mm.}
\label{fig:PIXIE_fts}
\end{figure}

PIXIE's science instrument is a polarizing FTS with 14 optical surfaces per left or right beam. A cartoon of the instrument is shown in Figure \ref{fig:PIXIE_fts}, and a complete description of the optical design is given by \cite{Kogutetal2011, Kogutetal2014}. Light incident on the left or right side of the instrument is routed into the FTS. Polarizer A defines the polarization basis of the instrument, transmitting one polarization and reflecting the orthogonal polarization. The polarization angle of the instrument will be measured on the ground and calibrated in flight by observing a well-known polarized sky source \citep[e.g.][]{Aumontetal2010}. Polarizer B is oriented at 45 degrees relative to polarizer A, mixing the beams. The moving mirror injects an optical phase delay in the two beams, which are subsequently combined by polarizer C and sorted by polarizer D. The light is then routed into the square receiver horns and onto the focal planes. Each focal plane consists of two polarization-sensitive bolometers mounted back-to-back with their polarization axes orthogonal to each other, allowing simultaneous measurements of both polarizations. 

The spacecraft will spin about a common co-pointed beam axis at $\sim$ 4 rpm, amplitude modulating linearly polarized signals at twice this frequency. Either beam can be blocked by a full-aperture blackbody calibrator which will be kept within a few mK of the CMB temperature. The calibrator enables absolute characterization of PIXIE's optical efficiency and independent, non-differential measurements of each beam. Explicit expressions for the signal measured by the instrument when the calibrator blocks one of the beams are given by \cite{Kogutetal2011,Kogutetal2014}. 

The FTS will be kept isothermal with the CMB to within a few mK, so to first order any sky photon absorbed by the instrument will be replaced by an indistinguishable photon emitted by the instrument. We will actively control and asynchronously modulate the temperature of each surface in the FTS about the CMB temperature. The modulations will have periods of a few hours and will nearly be mutually orthogonal. Their amplitudes will vary, but never exceed a few mK. This enables significant control over systematic effects.

Deployed from low earth orbit at an altitude of 660 km, PIXIE will measure the same sky pixel $\sim250$ times every 34 hours, when, at the ecliptic, the pointing precesses to the next pixel. This enables us to take sums and differences of measured signals under the reasonable assumption that the sky signal is constant, playing an important role in mitigating systematic effects.

The low frequency response of the instrument is inhibited by the etendu. Specifically there is a low frequency cutoff at 15 GHz imposed by the physical size of the detector. This is reinforced by the maximum scan length which limits the resolution to $\sim$ 15 GHz and thus all of the power below $\sim$ 30 GHz is pushed into inaccessible bins in the Fourier transform.

The highest frequency is limited by the spacing on the wire grids. A filter (or series of filters) will be included to limit the input and hence noise from frequencies greater than $\sim$ 5 THz. These are further limited by the finite response time of the bolometers.

The response of the instrument will not be uniform over all frequencies but these are calibrated by observing the changing temperature of the calibrator.

\subsection{Jones matrix method}\label{sub:Jones-matrix-method}

The Jones matrix method \citep{Jones1941} is used to model the PIXIE FTS. Mathematically equivalent to either the Mueller matrix method \citep{Mueller1943} or the coherency matrix method \citep{Wolf1959}, the Jones method assigns a 2$\times$2 matrix operator $\mathbf{J}_{i}$ to the $i^{th}$ optical element in the signal path. Radiation incident on the $i^{th}$ element is represented by the vector $E_{i-1}$, and the polarization state $E_{i}$ after interaction with the optical element is given by 

\begin{equation}
E_{i}=\mathbf{J}_{i}E_{i-1}.
\end{equation}

Now we define the Jones operators that correspond to the ideal instrument's optical components, specifically mirrors, wire grid polarizers, and the moving mirror assembly. It is possible to define Jones operators in either the reference frame of the instrument or the reference frame of the radiation; we choose the latter. This choice does not affect the form of the resulting power expressions.

Reflections off a mirror flip the signs of the incident electric fields, but conserve power. Thus the corresponding Jones reflection operator is

\begin{equation}
\mathbf{J}_{r}=\left[\begin{array}{cc}
-1 & 0\\
0 & -1
\end{array}\right],
\end{equation}
where the subscript $r$ refers to reflections.

The wire grid polarizers transmit one polarization state while reflecting the orthogonal state. Thus each polarizer requires a Jones operator for both transmission and reflection. For a polarizer oriented such that $\hat{x}$ polarization is transmitted and $\hat{y}$ polarization is reflected, the corresponding Jones operators are

\begin{equation}
\begin{aligned}
\mathbf{J}_{t}&=\left[\begin{array}{cc}
1 & 0\\
0 & 0
\end{array}\right],\\ 
\mathbf{J}_{r}&=\left[\begin{array}{cc}
0 & 0\\
0 & -1
\end{array}\right],
\end{aligned}
\end{equation}
where the subscripts $t$ and $r$ refer to transmission and reflection, respectively. When the polarizer is instead oriented at some angle $\theta$ about the $\hat{z}$-axis, which is the direction of light propagation, then the Jones operators $\mathbf{J}_{t}\left(\theta\right)$ and $\mathbf{J}_{r}\left(\theta\right)$ become

\begin{equation}
\begin{aligned}
\mathbf{J}_{t}\left(\theta\right)&=\mathbf{R}_{Z}\left(\theta\right)\mathbf{J}_{t}\mathbf{R}_{Z}\left(\theta\right)^{\dagger},\\
\mathbf{J}_{r}\left(\theta\right)&=\mathbf{R}_{Z}\left(\theta\right)\mathbf{J}_{r}\mathbf{R}_{Z}\left(\theta\right)^{\dagger},
\end{aligned}
\end{equation}
where $\mathbf{R}_{Z}\left(\theta\right)$ is the rotation operator about the $\hat{z}$-axis.

Reflections off the moving mirror assembly insert phase delays and flip the sign of the $\hat{y}$ polarization. The corresponding Jones operators are

\begin{equation}
\begin{aligned}
\mathbf{J}_{r}^{L}&=\left[\begin{array}{cc}
\exp\left(\frac{2i\nu z}{c}\right) & 0\\
0 & -\exp\left(\frac{2i\nu z}{c}\right)
\end{array}\right],\\
\mathbf{J}_{r}^{R}&=\left[\begin{array}{cc}
\exp\left(\frac{-2i\nu z}{c}\right) & 0\\
0 & -\exp\left(\frac{-2i\nu z}{c}\right)
\end{array}\right],
\end{aligned}
\end{equation}
where $\nu$ is the frequency of light, $z$ is the moving mirror position, and $c$ is the speed of light. The superscripts $L$ and $R$ indicate whether light is incident on the mirror assembly from the left or right side, respectively.

\subsection{Calculated signal}\label{sub:Ideal}

Light incident on the left side of the FTS is represented by $E^{L}=\mathscr{A}\hat{x}+\mathscr{B}\hat{y}$, and light incident on the right side is represented by $E^{R}=\mathscr{C}\hat{x}+\mathscr{D}\hat{y}$. The power measured by a single detector is then\footnote{In general we only give the power measured by the left side $\hat{x}$ detector, but similar expressions for signals measured by the other detectors are both available \citep{Kogutetal2011,Kogutetal2014} and readily worked out.}:

\begin{equation}\label{eq:Plx_ideal}
\mathbf{P}_{x}^{L}=\frac{1}{2}\int\left(\mathscr{B}^{2}+\mathscr{C}^{2}\right)+\left(\mathscr{C}^{2}-\mathscr{B}^{2}\right)\cos\left(\frac{4\nu z}{c}\right)d\nu,
\end{equation}
where $\nu$ is the frequency of light, $c$ is the speed of light, $z$ is the moving mirror position, and the superscript $L$ and subscript $x$ indicate that the power is measured by the left side $\hat{x}$ detector.

This expression contains a DC term and a term modulated by the movement of the moving mirror assembly. The latter represents our measured interferogram or fringe pattern, and it is proportional to the Fourier transform of the incident difference spectrum. Taking the inverse Fourier transform of the measured power gives the incident spectrum $\mathbf{S}_{x}^{L}\left(\nu\right)$. Up to some constant, it is given by

\begin{equation}\label{eq:Slx_ideal}
\mathbf{S}_{x}^{L}\left(\nu\right)=\mathscr{C}_{\nu}^{2}-\mathscr{B}_{\nu}^{2},
\end{equation}
where the subscript $\nu$ indicates that we are working in the frequency domain.

The spectrum given by Equation \ref{eq:Slx_ideal} is equal to the Stokes $Q$ parameter in instrument-fixed coordinates. This highlights the most critical aspect of PIXIE: the instrument is sensitive only to polarized sources. If the sky were unpolarized, PIXIE would measure no fringe pattern. Instead it only sees the difference spectrum between orthogonal polarizations of light incident on the two sides of the instrument. The symmetry of the PIXIE FTS ensures that nearly all systematic error terms are also proportional to differences.

\section{emission errors}\label{sec:electromagnetic-error}

In this section we review emission errors and see how associated systematic error terms propagate, are identified, and are corrected. Only non-ideal optical components upstream of polarizer B can yield error terms that are modulated by mirror movement. The moving mirror assembly will inject no phase delay in thermal photons emitted from polarizer B and beyond; therefore such non-ideal components will only serve to attenuate the sky signal. 

To demonstrate how non-ideal components lead to modulated error terms, we consider the cases of emissive primary mirrors, emissive grids on polarizer A, and emissive non-optical surfaces, specifically the frames around the transfer mirrors. Emissive folding flats, secondary mirrors, and first transfer mirrors will give rise to systematic error terms identical in form to those from emissive primary mirrors, therefore we do not provide their derivations in this work. 

To demonstrate how non-ideal components lead to attenuation of the sky signal, we model polarizer B's grids as emissive.

\subsection{Review of theory}\label{sub:EE_theory}

According to Kirchhoff's law of thermal radiation, the emissive power of a body is given by the product of its absorptance and the Planck formula:
\begin{equation}
\mathbf{E_{\nu}}=\alpha_{\nu}\mathbf{B}_{\nu,T},
\end{equation}
where $\mathbf{E_{\nu}}$ is the emission power spectrum, $\alpha_{\nu}$ is the frequency-dependent absorptance spectrum (unitless), and $\mathbf{B}_{\nu,T}$ is the Planck spectrum with a temperature $T$. 

For sufficiently thick optical surfaces, where power that is not reflected by the surface is absorbed, the reflectance $\rho_{\nu}$ is related to the absorptance by

\begin{equation}
\rho_{\nu}=1-\alpha_{\nu}.
\end{equation}
 
At sufficiently low frequencies (e.g. the cutoff for gold is 3.5 THz), the reflectance spectrum of metals is given by

\begin{equation}
\rho_{\nu}=1-2\sqrt{\frac{\nu}{\sigma_{0}}},
\end{equation}
where $\nu$ is the frequency of radiation, and $\sigma_{0}$ is the DC conductivity \citep{BornandWolf1999}. As a result, the absorptance spectrum is given by

\begin{equation}\label{eq:alpha}
\alpha_{\nu}=2\sqrt{\frac{\nu}{\sigma_{0}}}.
\end{equation}
 
We define the absorption coefficient as $a_{\nu}$, then:

\begin{equation}
a_{\nu}a_{\nu}^{\star}=\alpha_{\nu}.
\end{equation}
As a result, $a_{\nu}$ is given by

\begin{equation}
a_{\nu}=\sqrt{\alpha_{\nu}}\exp\left[i\phi_{a}\right],
\end{equation}
where $\phi_{a}$ is a phase factor.

Similarly, for the reflection coefficient $r_{\nu}$:

\begin{equation}
r_{\nu}r_{\nu}^{\star}=1-\alpha_{\nu}.
\end{equation}
Then $r_{\nu}$ is given by

\begin{equation}\label{eq:R coeff-1}
r_{\nu}=\sqrt{1-a_{\nu}a_{\nu}^{\star}}\exp\left[i\phi_{r}\right],
\end{equation}
where $\phi_{r}$ is a phase factor. 

These expressions show up in modified Jones operators that describe the non-ideal instrument. We include the subscript $\nu$ in the following analyses only when we are working in the frequency domain.

\subsection{Primary mirrors with non-zero emissivity}\label{sub:Primary_nze}

Here we investigate degradation due to emission mismatches from the primary mirrors.  Geometric mismatches leading to beam errors are explicitly left for an optical treatment.  In principle emission can be easily calculated from the conductivity of the material (Equation \ref{eq:alpha}), however, it is very sensitive to the details of the surface, cleanliness, surface roughness, temperature, oxidation, {\sl et cetera}. So we will leave the actual emissivity as a parameter in the following derivations.

\subsubsection{Modified operators}

When the primary mirrors are modeled to exhibit non-zero emissivity, we define two new reflection operators that correspond to reflections off the left and right primary mirrors, as indicated by the superscripts $L$ and $R$:

\begin{equation}\label{eq:Rm}
\begin{aligned}
\mathbf{J}_{r}^{L}&=\left[\begin{array}{cc}
-\left|r_{x}^{L}\right|\exp i\phi_{rx}^{L} & 0\\
0 & -\left|r_{y}^{L}\right|\exp i\phi_{ry}^{L}
\end{array}\right],\\
 \mathbf{J}_{r}^{R}&=\left[\begin{array}{cc}
-\left|r_{x}^{R}\right|\exp i\phi_{rx}^{R} & 0\\
0 & -\left|r_{y}^{R}\right|\exp i\phi_{ry}^{R}
\end{array}\right],
\end{aligned}
\end{equation}
where $r_{x}$ and $r_{y}$ are the reflection coefficients in the $\hat{x}$ and $\hat{y}$ directions, and $\phi_{rx}$ and $\phi_{ry}$ are phases between incident and reflected radiation in the $\hat{x}$ and $\hat{y}$ directions, respectively. 

The left and right primary mirrors also emit radiation whose electric fields are described by:

\begin{equation}\label{eq:Em}
\begin{aligned}
E_{M}^{L}&=\left[\begin{array}{c}
\frac{1}{\sqrt{2}}\left|a_{x}^{L}\right|\sqrt{\mathbf{B}_{\nu,T}}\exp i\phi_{ex}^{L}\\
\frac{1}{\sqrt{2}}\left|a_{y}^{L}\right|\sqrt{\mathbf{B}_{\nu,T}}\exp i\phi_{ey}^{L}
\end{array}\right],\\
 E_{M}^{R}&=\left[\begin{array}{c}
\frac{1}{\sqrt{2}}\left|a_{x}^{R}\right|\sqrt{\mathbf{B}_{\nu,T}}\exp i\phi_{ex}^{R}\\
\frac{1}{\sqrt{2}}\left|a_{y}^{R}\right|\sqrt{\mathbf{B}_{\nu,T}}\exp i\phi_{ey}^{R}
\end{array}\right],
\end{aligned}
\end{equation}
where $a_{x}$ and $a_{y}$ are the absorption coefficients in the $\hat{x}$ and $\hat{y}$-directions, $\phi_{ex}$ and $\phi_{ey}$ are phase factors between incident and emitted radiation in the $\hat{x}$ and $\hat{y}$ directions, and the factor of $1/\sqrt{2}$ normalizes the power such that $\left|E_{M}\right|^{2}=\mathbf{B}_{\nu,T}$. The subscript $M$ indicates the emission is from a mirror. The absorption and reflection coefficients in Equations \ref{eq:Rm} and \ref{eq:Em} are related by Equation \ref{eq:R coeff-1}.

\subsubsection{Calculated signal}\label{sub:Sims_primary}

Treating all optical surfaces except the primary mirrors as ideal, we get the following fringe pattern from the left side $\hat{x}$ detector:

\begin{multline}\label{eq:Plx_primary emission}
\tilde{\mathbf{P}}_{x}^{L}=\frac{1}{2}\int\bigg(\mathscr{C}^{2}\left(1-\alpha_{x}^{R}\right)-\mathscr{B}^{2}\left(1-\alpha_{y}^{L}\right)\\+\frac{1}{2}\left(\alpha_{x}^{R}-\alpha_{y}^{L}\right)\mathbf{B}_{\nu,T}\bigg)\cos\left(\frac{4\nu z}{c}\right)d\nu,
\end{multline}
where $\alpha_{x}$ and $\alpha_{y}$ are the absorptances of the primary mirror along the $\hat{x}$ and $\hat{y}$ directions, and the tilde indicates that this is an actual, not ideal, quantity. None of the phase factors present in Equations \ref{eq:Rm} and \ref{eq:Em} show up in the fringe pattern expression because the phases are uncorrelated quantities; consequently photons emitted by different optical components do not interfere with each other.

$\tilde{\mathbf{P}}_{x}^{L}$ is proportional to the Fourier transform of the radiation incident on the detector, so taking the inverse Fourier transform yields the spectrum ${\bf \tilde{S}}_x^L\left(\nu\right)$. Expressed as the sum of the spectrum measured by the ideal instrument and an error term $\epsilon_x^L\left(\nu\right)$, it is given by

\begin{equation}
\mathbf{\tilde{S}}_{x}^{L}\left(\nu\right)=\mathscr{C}_{\nu}^{2}-\mathscr{B}_{\nu}^{2}+\epsilon_{x}^{L}\left(\nu\right).
\end{equation}
The error term is given by

\begin{equation}\label{eq:Elx_primary emission}
\epsilon_{x}^{L}\left(\nu\right)=\Delta\alpha_{\nu}\left(-\mathbf{B}_{\nu,T}+\mathscr{B}_{\nu}^{2}+\mathscr{C}_{\nu}^{2}\right),
\end{equation}
where $\Delta\alpha_{\nu}$ is the difference in absorptance between the left and right primary mirrors. 

The error term shows up in second order and is proportional to the {\sl difference} in absorptance between the left and right primary mirrors, multiplied by the {\sl difference} in temperature between the instrument and the sky. The difference in absorptance will be of order $10^{-4}$. This assumes that the average absorptance of the mirrors is of order $10^{-2}$ and they are matched to a part in $10^{2}$. Both these constraints are easily met. Well-polished aluminum can demonstrate emissivity of order $10^{-3}$ at PIXIE frequencies \citep{Bocketal1995}, and both mirrors will be made from the same aluminum stock. The component of the error term given by the difference spectrum between the sky and the instrument ({\sl i.e.} the quantity $\mathbf{-B}_{\nu,T}+\mathscr{B}_{\nu}^{2}+\mathscr{C}_{\nu}^{2}$ in Equation \ref{eq:Elx_primary emission}) will be a few mK. Uncorrected, the error signal from emissive mirrors will be hundreds of nK, larger than the 100 nK {\sl B}-mode polarization signal. The PIXIE design enables identification and correction of these errors down to the sub-nK level. 

\subsubsection{Identifying and correcting error}\label{sub:Error_primary}

Here we focus on identifying and correcting the error given by Equation \ref{eq:Elx_primary emission}. We take advantage of several important elements of PIXIE, including the ability to keep the instrument isothermal with the CMB to within a few mK, the ability to actively control and modulate the temperature of each optical component, and the ability to observe the same sky pixel multiple times on short timescales. In the following derivations, we mitigate error only by performing linear operations on measurements performed by a single detector, avoiding any assumptions of uniformity between detector pairs. In practice, however, it is also possible to mitigate error by comparing signals measured by detector pairs.

\paragraph{Cold optics}

PIXIE's optics will be isothermal with the sky to within a few mK. For a given differential absorptance $\Delta\alpha_{\nu}$, the error will be determined by the temperature difference between the optics and the sky, denoted $\delta T$. Explicitly, if the mirrors are at some temperature $T=T_{0}+\delta T$, where $T_{0}$ is the CMB temperature, then the error given by Equation \ref{eq:Elx_primary emission} becomes

\begin{equation}\label{eq:Elx_primary_corr1}
\epsilon_{x}^{L}\left(\nu\right)=\Delta\alpha_{\nu}\left(-\mathbf{B}_{\nu,T_{0}+\delta T}+\mathbf{B}_{\nu,T_{0}}\right),
\end{equation}
where we used $\mathbf{B}_{\nu,T_{0}}=\mathscr{B}_{\nu}^{2}+\mathscr{C}_{\nu}^{2}$.

Since $\delta T$ is small relative to $T_{0}$, we express the quantity $\mathbf{B}_{\nu,T_{0}+\delta T}$ as a Taylor series:

\begin{equation}
\mathbf{B}_{\nu,T_{0}+\delta T}=\mathbf{B}_{\nu,T_{0}}+\mathbf{B}_{\nu,T_{0}}^{\prime}\delta T+\Theta\left(\delta T^{2}\right),
\end{equation}
where

\begin{equation}\label{eq:B_prime}
\mathbf{B}_{\nu,T_{0}}^{\prime}=\frac{\partial\mathbf{B}_{\nu,T}}{\partial T}\bigg|_{T=T_{0}}.
\end{equation}

Then the error reduces to

\begin{equation}
\epsilon_{x}^{L}\left(\nu\right)=-\Delta\alpha_{\nu}\mathbf{B}_{\nu,T_{0}}^{\prime}\delta T.
\end{equation}

Thus error scales linearly with the temperature difference $\delta T$ between the sky and the mirrors. Over the course of the mission, we will match each warm temperature $T=T_{0}+\delta T$ with a cooler temperature $T=T_{0}-\delta T$. As a result, the error will reduce to a higher order term proportional to errors in controlling $\delta T$, contributing not to mean measurements, but rather to measurement variance.

\paragraph{Mirror temperature modulation}

The primary mirrors will be kept close to the CMB temperature $T_{0}$, but we actively modulate and control the temperature difference $\delta T$ between the mirrors and the sky. Keeping terms through second order, the emission spectrum of the mirrors is

\begin{multline}\label{eq: Mirror emissivity}
\Delta\alpha_{\nu}\mathbf{B}_{\nu,T}=\Delta\alpha_{\nu}\bigg(\mathbf{B}_{\nu,T_{0}}+\mathbf{B}_{\nu,T_{0}}^{\prime}\left(T-T_{0}\right)\\
+\frac{1}{2}\mathbf{B}_{\nu,T_{0}}^{\prime\prime}\left(T-T_{0}\right)^{2}+\Theta\left(T-T_{0}\right)^{3}\bigg).
\end{multline}

If the mirror is at $T=T_{0}+\delta T$, the emission spectrum becomes

\begin{multline}
\Delta\alpha_{\nu}\mathbf{B}_{\nu,T_{0}+\delta T}=\Delta\alpha_{\nu}\bigg(\mathbf{B}_{\nu,T_{0}}+\mathbf{B}_{\nu,T_{0}}^{\prime}\delta T\\
+\frac{1}{2}\mathbf{B}_{\nu,T_{0}}^{\prime\prime}\delta T^{2}+\Theta\left(\delta T\right)^{3}\bigg).
\end{multline}

Similarly for $T=T_{0}-\delta T$, the emission spectrum becomes

\begin{multline}
\Delta\alpha_{\nu}\mathbf{B}_{\nu,T_{0}-\delta T}=\Delta\alpha_{\nu}\bigg(\mathbf{B}_{\nu,T_{0}}-\mathbf{B}_{\nu,T_{0}}^{\prime}\delta T\\+\frac{1}{2}\mathbf{B}_{\nu,T_{0}}^{\prime\prime}\delta T^{2}-\Theta\left(\delta T\right)^{3}\bigg).
\end{multline}
 
Now imagine that during the observation of one sky pixel, say at time $t=t_{1}$, the primary mirrors are held at $T=T_{0}+\delta T$. During a subsequent observation of the same sky pixel, at time $t=t_{2}$, the mirrors are kept at $T=T_{0}-\delta T$. If we difference the power measured by a single detector at the two observation times, the sky signals vanish, leaving

\begin{multline}
\tilde{\mathbf{S}}_{x}^{L}\left(\nu\right)\bigg|_{t=t_{1}}-\tilde{\mathbf{S}}_{x}^{L}\left(\nu\right)\bigg|_{t=t_{2}}\\=-2\Delta\alpha_{\nu}\,\mathbf{B}_{\nu,T_{0}}^{\prime}\delta T+\Theta\left(\delta T^{3}\right).
\end{multline}
Since $\mathbf{B}_{\nu,T_0}^\prime$ and $\delta T$ are known, differencing gives a direct measure of $\Delta\alpha_{\nu}$ to nK-scale precision.

With $\Delta\alpha_{\nu}$ measured, consider summing the power measured at the two observation times:

\begin{multline}\label{eq:Slx_primary_summing}
\tilde{\mathbf{S}}_{x}^{L}\left(\nu\right)\bigg|_{t=t_{1}}+\tilde{\mathbf{S}}_{x}^{L}\left(\nu\right)\bigg|_{t=t_{2}}\\=2\mathscr{C}_{\nu}^{2}-2\mathscr{B}_{\nu}^{2}+\Sigma\epsilon_{x}^{L}\left(\nu\right),
\end{multline}
where

\begin{equation}\label{eq:Elx_primary_summing}
\Sigma\epsilon_{x}^{L}\left(\nu\right)=\Delta\alpha_{\nu}\mathbf{B}_{\nu,T_{0}}^{\prime\prime}\delta T^{2}.
\end{equation}

If the temperature excursions $\delta T$ are kept at the few-mK level, then the residual error given by Equation \ref{eq:Elx_primary_summing} will be $\sim10^{-10}$ K, much smaller than the expected {\sl B}-mode signal. It is easily confirmed that the operations that mitigate the left side $\hat{x}$ error to the sub-nK level simultaneously apply to the error measured by the other three detectors.

\begin{figure}[t]
\centering
\includegraphics[width=1\columnwidth]{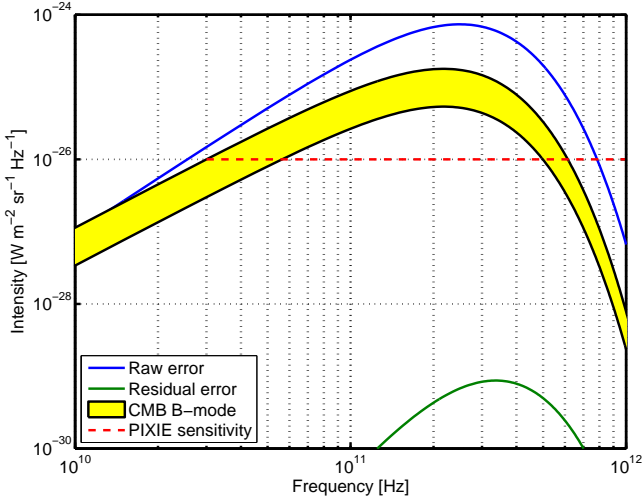}
\caption{Error due to emissive primary mirrors. Both the CMB signal and the raw error are proportional to $\mathbf{B}_{\nu,T_{0}}^{\prime}$ (Equation \ref{eq:B_prime}), while the residual error is proportional to $\mathbf{B}_{\nu,T_{0}}^{\prime\prime}$. They can therefore be distinguished based on their spectral content. In addition the error terms exhibit the $\nu^{1/2}$ dependence of the absorptance spectrum. We have conservatively assumed that we will know both $\delta T$ and $\Delta\alpha_{\nu}$ to within $\sim1\%$. The {\sl B}-mode band shows amplitudes for $0.01<r<0.1$.}
\label{fig: Error_emissive primary}
\end{figure}

\subsubsection{Conclusion}

This section shows how emissive primary mirrors affect measured signals. Uncorrected error signals in the raw time stream are hundreds of nK. But PIXIE's design enables significant control over these error signals. By performing simple operations - modulating the temperature of the primary mirrors and taking sums and differences of power measured by single detectors - the errors can be suppressed to a level that is negligible compared to PIXIE noise (see Figure \ref{fig: Error_emissive primary}).

Over the course of the mission, however, the error will become even smaller than is represented by Equation \ref{eq:Elx_primary_summing}. We will take linear combinations of signals from not just two measurements of the same pixel of sky, but repeated measurements of all sky pixels. This enables knowledge of $\delta T$ and $\Delta\alpha_{\nu}$ with ever increasing precision. As a result, we will subtract away error of the form given by Equation \ref{eq:Elx_primary_summing}, leaving some higher order term whose magnitude is $<1\%$ of the error given by Equation \ref{eq:Elx_primary_summing}.

\subsection{Polarizer grid A with non-zero emissivity}\label{sub:Non-ideal-polarizing-grid}

The PIXIE polarizing grids will be grids of fine wire stretched over a stiff frame.  We have constructed grids of 12 $\mu$m diameter tungsten wire with 30 $\mu$m spacing with $\le$ 5 $\mu$m rms error in spacing and flatness over $\sim$ 100 mm diameters. For frequencies less than 500 GHz ({\sl i.e.} most of the CMB spectrum) these are essentially perfect \citep{Chussetal2012}. By employing free-standing wire grid polarizers and keeping the FTS isothermal with the sky, we avoid some of the beam splitter systematics others have observed \citep[e.g.][and references therein]{Spenceretal2011}. Thus our treatment of systematics stemming from the polarizing grids is limited to emission and absorption (this section) and misalignments (Section \ref{sub:misal_pol}).

When a polarizing grid is modeled to exhibit non-zero emissivity, its reflection and transmission operators change, and it also emits thermal radiation. We treat polarizer A as emissive, demonstrating how errors appear in the measured fringe pattern. Later we consider the case of an emissive polarizer B, which only attenuates the sky signal.

\subsubsection{Modified operators}

There are four relevant operators which act on transmitted or reflected light incident from the left or right side of an emissive polarizer A. The transmission and reflection operators that act on radiation incident from the left are given by

\begin{equation}\label{eq:Operators_LHS_grid}
\begin{aligned}
\mathbf{J}_{t}^{L}&=\left[\begin{array}{cc}
1 & 0\\
0 & 0
\end{array}\right],\\
\mathbf{J}_{r}^{L}&=\left[\begin{array}{cc}
0 & 0\\
0 & -\left|r\right|\exp i\phi_{r}^{L}
\end{array}\right],
\end{aligned}
\end{equation}
where $\left|r\right|$ is the reflection coefficient of the grid and $\phi_{r}^{L}$ is the phase between incident and reflected light off the grid. The transmission operator does not change since light whose electric field is orthogonal to the wire grids is not sensitive to emissive wires.

The equivalent operators that act on radiation incident from the right are given by

\begin{equation}\label{eq:Operators_RHS_grid}
\begin{aligned}
\mathbf{J}_{t}^{R}&=\left[\begin{array}{cc}
1 & 0\\
0 & 0
\end{array}\right],\\
\mathbf{J}_{r}^{R}&=\left[\begin{array}{cc}
0 & 0\\
0 & -\left|r\right|\exp i\phi_{r}^{R}
\end{array}\right],
\end{aligned}
\end{equation}
where $\left|r\right|$ is the reflection coefficient and $\phi_{r}^{R}$ is the associated phase factor. We assume here that both sides of the grid have the same reflection coefficient. This is not strictly the case, but any differences between the two will be small. For details see \cite{Chussetal2012}.

The grids also emit thermal radiation along the same direction as reflections. The electric fields from this radiation are represented by

\begin{equation}\label{eq:Emiss_grid}
\begin{aligned}
E_{P}^{L}&=\left[\begin{array}{c}
0\\
\frac{1}{2}\left|a\right|\sqrt{\mathbf{B}_{\nu,T}}\exp i\phi_{e}^{L}
\end{array}\right],\\
E_{P}^{R}&=\left[\begin{array}{c}
0\\
\frac{1}{2}\left|a\right|\sqrt{\mathbf{B}_{\nu,T}}\exp i\phi_{e}^{R}
\end{array}\right],
\end{aligned}
\end{equation}
where $\left|a\right|$ is the absorption coefficient of the grid, $\phi_{e}$ is the phase between incident and emitted radiation, and the factor of $1/2$ normalizes the power such that $\left|E_{P}^{L}\right|^{2}+\left|E_{P}^{R}\right|^{2}=\mathbf{B}_{\nu,T}/2$. The subscript $P$ indicates that this is emission from a polarizer.

\subsubsection{Calculated signal}

Treating all optical surfaces except polarizer A as ideal, we get the following fringe pattern:

\begin{multline}
\tilde{\mathbf{P}}_{x}^{L}=\frac{1}{2}\int\left(\mathscr{C}^{2}-\mathscr{B}^{2}\left(1-\alpha\right)-\frac{1}{2}\alpha\mathbf{B}_{\nu,T}\right)\\ \times\cos\left(\frac{4\nu z}{c}\right)d\nu,
\end{multline}
where $\alpha$ is the absorptance of the grids. We take the inverse Fourier transform and express the resulting spectrum as the sum of the ideal instrument's measurement and an error term $\epsilon_x^L\left(\nu\right)$. Up to some constant, the spectrum is

\begin{equation}\label{eq:Slx_grids}
\tilde{\mathbf{S}}_{x}^{L}\left(\nu\right)=\mathscr{C}_{\nu}^{2}-\mathscr{B}_{\nu}^{2}+\epsilon_{x}^{L}\left(\nu\right),
\end{equation}
and error term is given by
 
\begin{equation}\label{eq:Elx_grids}
\epsilon_{x}^{L}\left(\nu\right)=\alpha_{\nu}\left(\mathscr{B}_{\nu}^{2}-\frac{1}{2}\mathbf{B}_{\nu,T}\right).
\end{equation}

The error from an emissive polarizer A is similar to that from emissive mirrors (Equation \ref{eq:Elx_primary emission}), except it is sensitive to absolute absorptance instead of differential absorptance. This is the only emissive systematic error that is sensitive to absolute absorptance. This is because only one polarization is reflected by the grid and therefore subject to attenuation. The other polarization passes through unaware that the grid is emissive. Similarly, emission from the grid is only in one polarization.
	
Before performing any corrective actions, we estimate the raw magnitude of the error. The quantity in parenthesis will be proportional to the temperature difference between the sky and the grid (a few mK). The absorptance of the grids $\alpha_{\nu}$ will be $\sim10^{-2}$. Therefore the error will be of order $10$ $\mu$K.
 
\subsubsection{Identifying and correcting error}

To identify and correct the error given by Equation \ref{eq:Elx_grids}, we modulate the temperature of the wire grids by some $\delta T$ about the CMB temperature and observe the same sky pixel multiple times. Again the details will be presented for measurements by the left side $\hat{x}$ detector, but the techniques employed simultaneously apply to measurements made by the other three detectors.
 
\paragraph{Cold optics}

The wire grids will be kept isothermal with the sky to within a few mK. The quantity $\mathscr{B}_{\nu}^{2}$ will be a Planck spectrum at CMB temperature $T_{0}$, but at half the overall intensity since it represents only one polarization state. Then the error given by Equation \ref{eq:Elx_grids} reduces to
\
\begin{equation}\label{eq:elx_grids_cold}
\epsilon_{x}^{L}\left(\nu\right)=\alpha_{\nu}\left(\frac{1}{2}\mathbf{B}_{\nu,T_{0}}-\frac{1}{2}\mathbf{B}_{\nu,T}\right).
\end{equation}

Since the temperature of the wire grids will be close to that of the sky, we represent the emission spectrum of the grids as a Taylor series. Then the error becomes

\begin{equation}\label{eq:Elx_first}
\epsilon_{x}^{L}\left(\nu\right)=\frac{1}{2}\alpha_{\nu}\mathbf{B}_{\nu,T_{0}}^{\prime}\delta T.
\end{equation}

The error due to an emissive grid will scale linearly with the temperature difference between the grids and the sky. As with the primary mirrors, each warm temperature $T=T_{0}+\delta T$ will be matched with a cooler temperature $T=T_{0}-\delta T$. Then the error given by Equation \ref{eq:Elx_first} will ultimately reduce to a higher order term that depends on the uncertainty of $\delta T$.

\paragraph{Grid temperature modulation}

We mitigate the error due to an emissive grid by modulating the grid's temperature and taking linear combinations of measured signals. Assume that during the observation of a pixel of sky at time $t=t_{1}$, the emissive wire grid polarizer is at temperature $T=T_{0}+\delta T$, and that during a subsequent observation of the same sky pixel at time $t=t_{2}$, the polarizer is at temperature $T=T_{0}-\delta T$. Differencing the power measured at times $t_1$ and $t_2$ gives

\begin{equation}
\tilde{\mathbf{S}}_{x}^{L}\left(\nu\right)\bigg|_{t=t_{1}}-\tilde{\mathbf{S}}_{x}^{L}\left(\nu\right)\bigg|_{t=t_{2}}=-\alpha_{\nu}\mathbf{B}_{\nu,T_{0}}^{\prime}+\Theta\left(\delta T^{3}\right).
\end{equation}

$\mathbf{B}_{\nu,T_0}^\prime$ is known, so differencing gives a measure of the average absorptance $\alpha_{\nu}$ of the wire grids. This operation also gives us the ability to measure the differential absorptance between two sides of the same grid. Examining 
$\epsilon_{x}^{L}\left(\nu\right)$ and $\epsilon_{y}^{R}\left(\nu\right)$ gives the absorptance of the left side of the grid, and examining $\epsilon_{y}^{L}\left(\nu\right)$ and $\epsilon_{x}^{R}\left(\nu\right)$ gives the absorptance of the right side.

Summing the power measured at times $t_{1}$ and $t_{2}$ gives

\begin{equation}
\tilde{\mathbf{S}}_{x}^{L}\left(\nu\right)\bigg|_{t=t_{1}}+\tilde{\mathbf{S}}_{x}^{L}\left(\nu\right)\bigg|_{t=t_{2}}=2\mathscr{C}_{\nu}^{2}-2\mathscr{B}_{\nu}^{2}+\Sigma\epsilon_{lx}\left(\nu\right),
\end{equation}
where the error term is given by

\begin{equation}\label{eq:Sum_Elx_grid_corrected}
\Sigma\epsilon_{x}^{L}\left(\nu\right)=-\alpha_{\nu}\mathbf{B}_{\nu,T_{0}}^{\prime\prime}\delta T^{2}.
\end{equation}

The residual error given by Equation \ref{eq:Sum_Elx_grid_corrected} is at the nK level, and its spectrum will differ from that of the sky.

\subsubsection{Conclusion}

\begin{figure}[t]
\centering
\includegraphics[width=1\columnwidth]{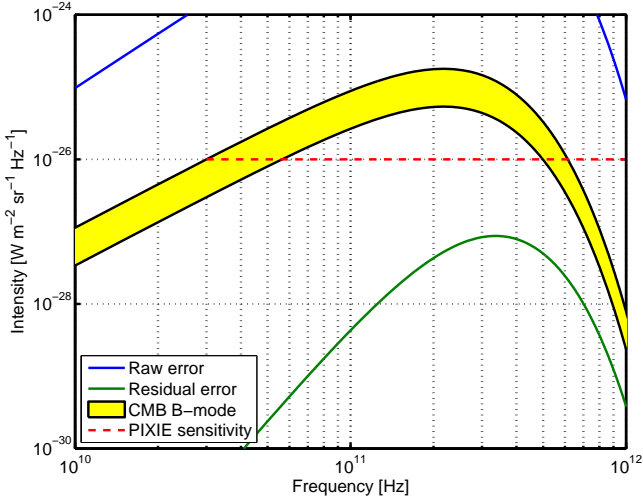}
\caption{Error due to an emissive polarizer A. Both the CMB signal and the raw error are proportional to $\mathbf{B}_{\nu,T_{0}}^{\prime}$, while the residual error is proportional to $\mathbf{B}_{\nu,T_{0}}^{\prime\prime}$. We have conservatively assumed that we will know both $\delta T$ and $\alpha_{\nu}$ to within $\sim1\%$, so the residual error is a factor of 100 smaller than expressed in Equation \ref{eq:Sum_Elx_grid_corrected}. If we instead take sums and differences of signals measured by different detectors, both the raw and residual error will be at least two orders of magnitude smaller, and only sensitive to differential, rather than absolute, absorptance. The {\sl B}-mode band shows amplitudes for $0.01<r<0.1$.\vspace{0.01 cm}}
\label{fig: Error_emissive polA}
\end{figure}

This section shows how error from an emissive polarizer A will affect the measured signals. Uncorrected, errors are of order $10$ $\mu$K. By modulating the temperature of the emissive grid and taking linear combinations of power measured at different times, these errors can be reduced to below 1 nK (see Figure \ref{fig: Error_emissive polA}). If we instead take sums and differences of signals measured by detector pairs, then both the raw and residual errors will be proportional to the differential absorptance between the two sides of the grid, taking the same form as error from emissive primary mirrors (Equations \ref{eq:Elx_primary emission} and \ref{eq:Elx_primary_summing}, respectively).  Their magnitudes will be at least two orders of magnitude smaller than in Figure \ref{fig: Error_emissive polA}.

\subsection{Polarizer B with non-zero emissivity}\label{sub:Polarizer-B}

In this section, we model polarizer B as exhibiting non-zero emissivity, demonstrating how non-ideal optical components downstream of the second transfer mirrors affect our measurements. Their effect is different from the previous two examples because thermal photons emitted by such surfaces are not split, mixed, and interfered, thus they contribute no modulated error signals. Instead, non-unity reflectivity only attenuates our sky signals, an effect that is measured with the calibrator.

\subsubsection{Modified operators}

Polarizer B is oriented at $45^{\circ}$ relative to polarizer A, so its non-ideal operators are derived by applying the rotation operator to polarizer A's modified operators. Then the transmission and reflection operators that act on radiation incident from the left are given by

\begin{equation}\label{eq:Operators_LHS_grid-1}
\begin{aligned}
\mathbf{J}_{t}^{L}&=\mathbf{R}_{Z}\left(45^{\circ}\right)\times\left[\begin{array}{cc}
1 & 0\\
0 & 0
\end{array}\right]\times\mathbf{R}_{Z}^{\dagger}\left(45^{\circ}\right),\\
\mathbf{J}_{r}^{L}&=\mathbf{R}_{Z}\left(45^{\circ}\right)\times\left[\begin{array}{cc}
0 & 0\\
0 & -\left|r\right|\exp i\phi_{r}^{L}
\end{array}\right]\times\mathbf{R}_{Z}^{\dagger}\left(45^{\circ}\right).
\end{aligned}
\end{equation}
 
The equivalent operators that act on radiation incident from the right are given by

\begin{equation}\label{eq:Operators_RHS_grid-1}
\begin{aligned}
\mathbf{J}_{t}^{R}&=\mathbf{R}_{Z}\left(45^{\circ}\right)\times\left[\begin{array}{cc}
1 & 0\\
0 & 0
\end{array}\right]\times\mathbf{R}_{Z}^{\dagger}\left(45^{\circ}\right),\\
\mathbf{J}_{r}^{R}&=\mathbf{R}_{Z}\left(45^{\circ}\right)\times\left[\begin{array}{cc}
0 & 0\\
0 & -\left|r\right|\exp i\phi_{r}^{R}
\end{array}\right]\times\mathbf{R}_{Z}^{\dagger}\left(45^{\circ}\right).
\end{aligned}
\end{equation}

The grids emit thermal radiation along the same direction as reflections. The electric fields from this radiation are given by

\begin{equation}\label{eq:Emiss_grid-1}
\begin{aligned}
E_{P}^{L}&=\mathbf{R}_{Z}\left(45^{\circ}\right)\times\left[\begin{array}{c}
0\\
\frac{1}{2}\left|a\right|\sqrt{\mathbf{B}_{\nu,T}}\exp i\phi_{e}^{L}
\end{array}\right],\\
E_{P}^{R}&=\mathbf{R}_{Z}\left(45^{\circ}\right)\times\left[\begin{array}{c}
0\\
\frac{1}{2}\left|a\right|\sqrt{\mathbf{B}_{\nu,T}}\exp i\phi_{e}^{R}
\end{array}\right].
\end{aligned}
\end{equation}
Again we set the absorption coefficient $\left|a\right|$ of the two sides of the grid to be equal.

\subsubsection{Calculated signal}

Treating all optical surfaces but polarizer B as ideal, we get the following fringe pattern:
 
\begin{equation}
\tilde{\mathbf{P}}_{x}^{L}=\frac{1}{2}\int\left|r\right|\left(\mathscr{C}^{2}-\mathscr{B}^{2}\right)\cos\left(\frac{4\nu z}{c}\right)d\nu,
\end{equation}
where $\left|r\right|$ is the reflection coefficient of the grid. We take the inverse Fourier transform to get the spectrum:

\begin{equation}\label{eq:Slx_grids-1}
\tilde{\mathbf{S}}_{x}^{L}\left(\nu\right)=\left|r\right|\left(\mathscr{C}_{\nu}^{2}-\mathscr{B}_{\nu}^{2}\right).
\end{equation}

No further steps need to be taken. The attenuation term $\left|r\right|$ is part of the instrument's optical efficiency, which is measured by the calibrator throughout the flight.

\subsubsection{Conclusion}

In this section we use the example of polarizer B to show how non-ideal optical components positioned downstream of the second transfer mirrors affect our measured signal. Unlike the examples of the primary mirrors and polarizer A, they contribute no systematic error, but instead only attenuate the sky signal. 
 
\subsection{Emission from non-optical components}

According to ray trace calculations, up to $\sim3\%$ of radiation incident on the detectors will not be from the sky or emitted by any optical components (such as mirrors or grids), but rather will be from emissive non-optical components, such as the transfer mirror frames. Equivalently, $\sim3\%$ of light incident on the instrument will be absorbed by non-optical components.  These frames and other components will be black, having emissivity approaching $1$. In this section we show that emission from surfaces upstream of polarizer B will result in residual error identical in form to that sourced by emissive primary mirrors.

\subsubsection{Emission from transfer mirror frames}\label{sub:no}

Radiation emitted by the transfer mirror frames is described by the following electric fields:

\begin{equation}\label{eq:Em-1}
\begin{aligned}
E_{F}^{L}&=\left[\begin{array}{c}
\frac{1}{\sqrt{2}}\sqrt{\kappa}\left|a_{x}^{L}\right|\sqrt{\mathbf{B}_{\nu,T}}\exp i\phi_{ex}^{L}\\
\frac{1}{\sqrt{2}}\sqrt{\kappa}\left|a_{y}^{L}\right|\sqrt{\mathbf{B}_{\nu,T}}\exp i\phi_{ey}^{L}
\end{array}\right],\\
E_{F}^{R}&=\left[\begin{array}{c}
\frac{1}{\sqrt{2}}\sqrt{\kappa}\left|a_{x}^{R}\right|\sqrt{\mathbf{B}_{\nu,T}}\exp i\phi_{ex}^{R}\\
\frac{1}{\sqrt{2}}\sqrt{\kappa}\left|a_{y}^{R}\right|\sqrt{\mathbf{B}_{\nu,T}}\exp i\phi_{ey}^{R}
\end{array}\right],
\end{aligned}
\end{equation}
where $\kappa$ is the fraction of power incident on the instrument that is absorbed by the frames and the subscript $F$ indicates that the radiation is coming from a transfer mirror frame.

The absorption coefficients present in Equation \ref{eq:Em-1} also act on incident radiation, such that a fraction $\kappa\alpha\sim0.03$ of power incident on the instrument is absorbed by the non-optical components.

\subsubsection{Calculated signal}

Treating all optical surfaces as ideal, and accounting for emission from the transfer mirror frames, we get the following fringe pattern:

\begin{multline}
\tilde{\mathbf{P}}_{x}^{L}=\frac{1}{2}\int\Big(\mathscr{C}^{2}\left(1-\kappa\alpha_x^L\right)-\mathscr{B}^{2}\left(1-\kappa\alpha_y^R\right)\\+\kappa\left(\alpha_{x}^{R}-\alpha_{y}^{L}\right)\mathbf{B}_{\nu,T}\Big)\cos\left(\frac{4\nu z}{c}\right)d\nu.
\end{multline}
Taking the inverse Fourier transform gives the spectrum:

\begin{equation}\label{eq:Slx_fram}
\tilde{\mathbf{S}}_{x}^{L}\left(\nu\right)=\mathscr{C}_{\nu}^{2}-\mathscr{B}_{\nu}^{2}+\epsilon_{x}^{L}\left(\nu\right).
\end{equation}
The associated error term is given by

\begin{equation}\label{eq:Elx_frame}
\epsilon_{x}^{L}\left(\nu\right)=\kappa_{\nu}\Delta\alpha_{\nu}\left(\mathbf{B}_{\nu,T}-\mathbf{B}_{\nu,T_0}\right),
\end{equation}
where $\Delta\alpha_{\nu}$ is the differential absorptance between the left and right frames.

Before taking any corrective actions, we estimate the magnitude of the error given by Equation \ref{eq:Elx_frame}. If $\sim3\%$ of all radiation incident on the detectors is emitted by non-optical components, then $\kappa_{\nu}\sim0.03$. The quantity $\Delta\alpha_{\nu}$ will be of order $10^{-2}$. The difference spectrum between the sky and the frames will be proportional to their temperature difference (a few mK). Therefore the error from emissive non-optical components will be hundreds of nK.
 
\subsubsection{Identifying and correcting error}

To identify and correct emissive frame errors, we actively control temperature of the frames and observe the same sky pixel multiple times. The operations shown apply to the signals measured by all detectors.

\paragraph{Cold optics}

For a given differential absorptance $\Delta\alpha_{\nu}$, the magnitude of the error is determined by the temperature difference between the frames and the sky. If the frames are at some temperature $T=T_{0}+\delta T$, then the error is

\begin{equation}\label{eq:Elx_frame_2}
\epsilon_{x}^{L}\left(\nu\right)=\kappa_{\nu}\Delta\alpha_{\nu}\mathbf{B}_{\nu,T_0}^{\prime}\delta T.
\end{equation}

The error scales linearly with the temperature difference $\delta T$ between the sky and the frames. Its spectrum will differ from polarized CMB signals since it is multiplied by $\kappa_{\nu}$, which exhibits the $\nu^{-1}$ dependence of diffraction off hard edges in the FTS. Since the average temperature of the frames over the whole mission will be very near $T_{0}$, the error will ultimately reduce to a higher order term proportional to uncertainty in $\delta T$.

\paragraph{Frame temperature modulation}

During the observation of one sky pixel, the transfer mirror frames are held at $T=T_{0}+\delta T$, and during a subsequent observation of the same sky pixel, the frames are kept at $T=T_{0}-\delta T$. Differencing the power measured at the two times gives:

\begin{multline}
\tilde{\mathbf{S}}_{x}^{L}\left(\nu\right)\bigg|_{t=t_{1}}-\tilde{\mathbf{S}}_{x}^{L}\left(\nu\right)\bigg|_{t=t_{2}}\\=2\kappa_{\nu}\Delta\alpha_{\nu}\Big(\mathbf{B}_{\nu,T_{0}}^{\prime}\delta T+\Theta\left(\delta T^{3}\right)\Big).
\end{multline}
Both $\mathbf{B}_{\nu,T_0}^\prime$ and $\delta T$ are well known, so differencing gives a measure of $\kappa_{\nu}\Delta\alpha_{\nu}$ to nK-scale precision.

With $\kappa_{\nu}\Delta\alpha_{\nu}$ measured, consider summing the measured powers:

\begin{equation}\label{eq:Sum_Slx-1}
\tilde{\mathbf{S}}_{x}^{L}\left(\nu\right)\bigg|_{t=t_{1}}+\tilde{\mathbf{S}}_{x}^{L}\left(\nu\right)\bigg|_{t=t_{2}}=2\mathscr{C}_{\nu}^{2}-2\mathscr{B}_{\nu}^{2}+\Sigma\epsilon_{x}^{L}\left(\nu\right),
\end{equation}
where

\begin{equation}\label{eq:Elx_primary_summing-1}
\Sigma\epsilon_{x}^{L}\left(\nu\right)=\kappa_{\nu}\Delta\alpha_{\nu}\mathbf{B}_{\nu,T_0}^{\prime\prime}\delta T^{2}.
\end{equation}

The residual error given by Equation \ref{eq:Elx_primary_summing-1} will be of order $10^{-10}$ K. 

\subsubsection{Conclusion}

\begin{figure}[t]
\centering
\includegraphics[width=1\columnwidth]{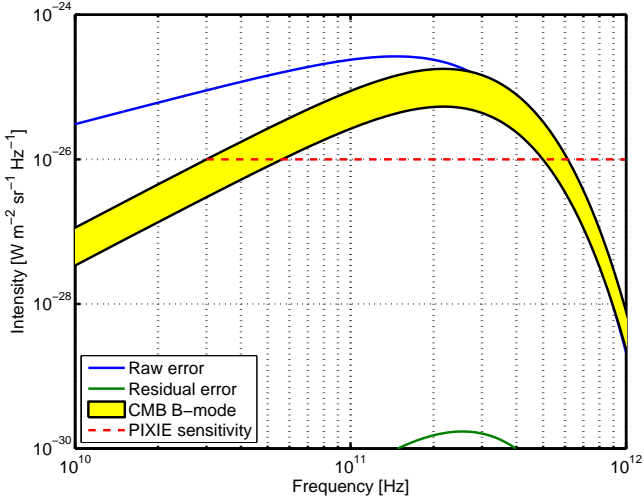}
\caption{Error due to emissive non-optical components. Both the CMB signal and the raw error are proportional to $\mathbf{B}_{\nu,T_{0}}^{\prime}$, while the residual error is proportional to $\mathbf{B}_{\nu,T_{0}}^{\prime\prime}$. Both the raw and residual error exhibit the $\nu^{-1}$ dependence of the diffraction coefficient $\kappa_{\nu}$. We have conservatively assumed that we will know both $\delta T$ and $\Delta\alpha_{\nu}$ to within $\sim1\%$, so the plotted corrected error is a factor of 100 smaller than expressed in Equation \ref{eq:Elx_primary_summing-1}. The {\sl B}-mode band shows amplitudes for $0.01<r<0.1$.}
\label{fig: Error_nonopt}
\end{figure}

In this section we show how systematic error terms from emissive non-optical components are identified and corrected. For reasons discussed in Section \ref{sub:Polarizer-B}, we are only concerned with thermal emission from components upstream of polarizer B. Uncorrected, such error is hundreds of nK. After performing corrective actions the residual error is well below $1$ nK (see Figure \ref{fig: Error_nonopt}).

\section{Geometric errors}\label{sub:Geometric-errors}

In this section we treat geometric errors. Of particular concern are geometric errors that lead to polarization leaks, which reduce the polarization sensitivity of the instrument. For example, a polarizing grid whose wires are not parallel allows transmission of light that should be reflected and reflection of light that should be transmitted. A polarizing grid that is misaligned by some angle about the direction of propagation of radiation has a similar effect. Broken grid wires also reduce the polarization sensitivity of the instrument. In the following we look specifically at misalignments of polarizers A and D, which are representative of any grid non-ideality that yields polarization leaks. 

We also consider transfer mirrors that are misaligned by some angle, but this case is best understood as a phase error in the Fourier transform. As such, it is treated in Section \ref{sec:MTMerr}. Misalignments of the primary mirrors, folding flats, and secondary mirrors lead to beam mismatches on the sky and are left to a separate work.

\subsection{Misalignment of polarizers A and D}\label{sub:misal_pol}

Here we model polarizers A and D to be misaligned by some small angle $\Delta\theta$ about the $\hat{z}$-axis. Since polarizers A and D define the polarization sensitivity, a common error only rotates the polarization axis. This can lead to power leakage from {\sl E}-modes to {\sl B}-modes, an error that is well treated in the literature \cite[e.g.][]{Huetal2003}, but it will not generate a polarized signal when the sky is unpolarized. Thus the concern is with differential errors. As such, polarizer A will be misaligned by $+\Delta\theta$, and polarizer D will be misaligned by $-\Delta\theta$. We will let all other optical surfaces, and their alignments, be ideal.

\subsubsection{Modified operators}

To account for small misalignments, we modify each polarizer's ideal reflection and transmission operators by applying the rotation operator $\mathbf{R}_Z\left(\pm\Delta\theta\right)$. Because the rotation angle $\Delta\theta$ will be small, the rotation operator for a misalignment of $\pm\Delta\theta$ is 

\begin{equation}
\mathbf{R}_{Z}\left(\Delta\theta\right)=\left[\begin{array}{cc}
1 & \mp\Delta\theta\\
\pm\Delta\theta & 1
\end{array}\right].
\end{equation}

Then polarizer A's transmission and reflection operators are:

\begin{equation}
\begin{aligned}
\mathbf{J}_{t}^{L}&=\mathbf{J}_{t}^{R}\\&=\mathbf{R}_{Z}\left(\Delta\theta\right)\times\left[\begin{array}{cc}
1 & 0\\
0 & 0
\end{array}\right]\times\mathbf{R}_{Z}^{\dagger}\left(\Delta\theta\right);
\end{aligned}
\end{equation}

\begin{equation}
\begin{aligned}
 \mathbf{J}_{r}^{L}&=\mathbf{J}_{r}^{R}\\&=\mathbf{R}_{Z}\left(\Delta\theta\right)\times\left[\begin{array}{cc}
0 & 0\\
0 & -1
\end{array}\right]\times\mathbf{R}_{Z}^{\dagger}\left(\Delta\theta\right),
\end{aligned}
\end{equation}
and the equivalent transmission and reflection operators for polarizer D are:

\begin{equation}
\begin{aligned}
\mathbf{J}_{t}^{L}&=\mathbf{J}_{t}^{R}\\&=\mathbf{R}_{Z}\left(-\Delta\theta\right)\times\left[\begin{array}{cc}
1 & 0\\
0 & 0
\end{array}\right]\times\mathbf{R}_{Z}^{\dagger}\left(-\Delta\theta\right);
\end{aligned}
\end{equation}

\begin{equation}
\begin{aligned}
 \mathbf{J}_{r}^{L}&=\mathbf{J}_{r}^{R}\\&=\mathbf{R}_{Z}\left(-\Delta\theta\right)\times\left[\begin{array}{cc}
0 & 0\\
0 & -1
\end{array}\right]\times\mathbf{R}_{Z}^{\dagger}\left(-\Delta\theta\right).
\end{aligned}
\end{equation}

\subsubsection{Calculated signal}

Treating all optical surfaces as having zero emissivity, and assuming all components are perfectly aligned except polarizers A and D, we get the following fringe pattern:

\begin{equation}
\tilde{\mathbf{P}}_{x}^{L}=\frac{1}{2}\int\left(\mathscr{C}^{2}-\mathscr{B}^{2}+4\Delta\theta\mathscr{BC}\right)\cos\left(\frac{4\nu z}{c}\right)d\nu.
\end{equation}
We take the inverse Fourier transform and express the spectrum as the sum of the ideal instrument's measurement and an error term:

\begin{equation}\label{eq:Slx_misalign}
\tilde{\mathbf{S}}_{x}^{L}\left(\nu\right)=\mathscr{C}_{\nu}^{2}-\mathscr{B}_{\nu}^{2}+\epsilon_{x}^{L}\left(\nu\right),
\end{equation}
where the error term is given by

\begin{equation}\label{eq:Elx_misalign_raw}
\epsilon_{x}^{L}\left(\nu\right)=4\Delta\theta\mathscr{B}_{\nu}\mathscr{C}_{\nu}.
\end{equation}

This error is proportional to the $\hat{y}$-polarized signal incident on the left side of the instrument multiplied by the $\hat{x}$-polarized signal incident on the right side. It will vanish when the calibrator is deployed since there can be no cross-polar response between mutually incoherent sources. Therefore the error is easily measured by differencing the signal measured with and without the calibrator. In addition this error vanishes in the absence of a polarized sky. To see this, it is useful to transform the error from instrument-fixed to sky-fixed coordinates (see Appendix \ref{app: Amplitude}). 

With both sides of the instrument open to the sky, the error due to misalignments in sky-fixed coordinates is

\begin{equation}\label{eq: ma_skyfixed}
\epsilon_{x}^{L}\left(\nu\right)=2\Delta\theta\left(U^{sky}\cos2\gamma+Q^{sky}\sin2\gamma\right),
\end{equation}
where $Q^{sky}$ and $U^{sky}$ are the Stokes parameters in sky-fixed coordinates and $\gamma$ is the spacecraft rotation angle.

As is clear from Equation \ref{eq: ma_skyfixed}, the error signal will vanish if the sky is unpolarized. Furthermore, it is a calibration term; it has the same form as the signal measured by the ideal instrument, except it is multiplied by the misalignment angle $\Delta\theta$. Misalignments of $\sim1/2^{\circ}$ will keep this term at the $1\%$ level. No further steps need to be taken to correct this error.

\subsection{Conclusion}

We show in this section how systematic error signals due to misaligned optical components originate, are identified, and are corrected. While we looked specifically at the case of rotated polarizers, the results and corrective actions applied are similar for any optical defect that causes a cross-polar response. Uncorrected, we get an error signal proportional to the misalignment angle, but it is easily measured and corrected by deploying the calibrator. By transforming the error from instrument-fixed to sky-fixed coordinates, we recognize that the error term is in fact a calibration term, so its presence reduces the signal-to-noise ratio by lowering our sensitivity to polarized signals, but it will not create a polarized signal from an unpolarized source.

If we model polarizer A to be both misaligned and emissive, we get an additional systematic error term similar in form to that given by Equation \ref{eq:Elx_grids}, except it will be multiplied by the misalignment angle $\Delta\theta$. As a result, it will be a factor of $\sim10^{3}$ smaller than the dominant emissive grid systematic, showing up at the sub-nK level without any corrections. The same operations that mitigate the dominant error given by Equation \ref{eq:Elx_grids} apply to this term.

\section{Mirror transport mechanism errors}\label{sec:MTMerr}

In this section we treat non-idealities in the mirror transport mechanism, which can lead to systematic error terms or contribute to the noise. If the zeroth sample of the interferogram is not at the null or zero phase position, where the path length of light traveling in either FTS beam is the same, there will be a corresponding phase error that asymmetrizes the interferogram. If the mirror stroke is not symmetric about the null position, the interferogram will also be asymmetric. Geometric misalignments of transfer mirrors lead to phase errors in the Fourier transform (this is not strictly a MTM error, but is best understood in this context). Uncertainties in the mirror position, which can relate to both timing and position errors, can lead to the presence of additional sidebands in the measured spectrum. We look at each of these errors in the general case, including a discussion of the steps that can be taken to mitigate them. Then we see how they affect the measurements.

\subsection{Phase errors}\label{sub:Phase-error}

\subsubsection{The ideal interferogram}

The theoretical treatment in the following closely follows the derivations presented by \cite{Connes1961}, \cite{Formanetal1966}, \cite{Sakaietal1968} and \cite{Bell1972}. An alternative approach is presented by \cite{Mertz1967}.

\begin{figure}[t]
\centering
\includegraphics[width=1\columnwidth]{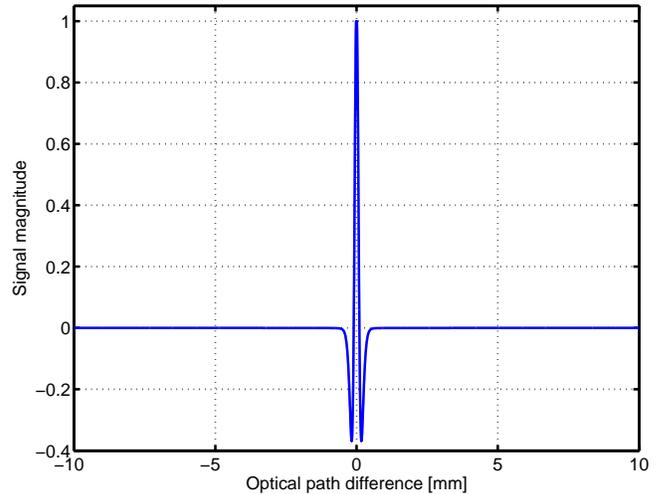}
\caption{The ideal two-sided interferogram showing the Fourier transform of the polarized CMB. Because the interferogram is symmetric about the null position, the spectrum can be recovered by performing the Fourier cosine transform on either half of the interferogram. The optical path difference $\ell$ is related to the frequency of incident radiation $\nu$ by the relation $\ell=c/\nu$. Similarly, the acoustic frequency $\omega$ of the mirror movement is related to the frequency of radiation by $\omega=4\nu v/c$, where $v$ is the moving mirror assembly's velocity. The CMB signal is largely confined to acoustic frequencies below 50 Hz and optical phase delays below 2 mm. The dust signal is constrained to acoustic frequencies below 200 Hz and optical phase delays below 50 $\mu$m.}
\label{fig:ifg_ideal}
\end{figure}

Consider the ideal case where both the interferogram and mirror stroke are symmetric about the null. In this case, the signal measured by a detector, say the left side $\hat{x}$ detector, will be given by

\begin{equation}\label{eq:interogram_ideal}
\mathbf{P}_{x}^{L}\left(\ell\right)=\int_{-\infty}^{\infty}\mathbf{S}_{x}^{L}\left(\nu\right)\cos\left(2\pi\nu\ell\right)d\nu,
\end{equation}
up to some constant factor. Here $\ell$ is the optical path difference between the two FTS beams, and $\nu$ is frequency expressed in wavenumber. The interferogram $\mathbf{P}_{x}^{L}\left(\ell\right)$ is a function of the MTM position $z\simeq\mbox{\ensuremath{\ell}/4}$, and the corresponding spectrum $\mathbf{S}_{x}^{L}\left(\nu\right)$ is a function of frequency $\nu$. The peak of the interferogram occurs at the null position $\ell=0$, and the interferogram is symmetric about this peak position (see Figure \ref{fig:ifg_ideal}). To recover the spectrum $\mathbf{S}_{x}^{L}\left(\nu\right)$ for the symmetric case, the full complex Fourier transform need not be applied. Rather we can apply just the even part (the Fourier cosine transform) to $\mathbf{P}_{x}^{L}\left(\ell\right)$. This operation gives, up to a constant factor,

\begin{equation}\label{eq:spectrum_ideal}
\mathbf{S}_{x}^{L}\left(\nu\right)=\int_{-\infty}^{\infty}\mathbf{P}_{x}^{L}\left(\ell\right)\cos\left(2\pi\nu\ell\right)d\ell.
\end{equation}
By symmetry, Equation \ref{eq:spectrum_ideal} can also be written as
\begin{equation}
\mathbf{S}_{x}^{L}\left(\nu\right)=2\int_{0}^{\infty}\mathbf{P}_{x}^{L}\left(\ell\right)\cos\left(2\pi\nu\ell\right)d\ell.
\end{equation}
The incident spectrum $\mathbf{S}_{x}^{L}\left(\nu\right)$ can be recovered independently by taking the Fourier cosine transform of either half of the full two-sided interferogram.

Equation \ref{eq:spectrum_ideal} assumes that the mirror throw is infinite. This will not be the case, so we need to include an apodization function $A\left(\ell\right)$ that takes into account the finite throw of the mirror. For PIXIE, the maximum mirror throw will be $\pm2.6$ mm, corresponding to a total optical path difference $L=\pm10$ mm. We will sample the interferogram at 1 kHz, and a complete mirror stroke from one extreme to the other takes 1 second. Assuming that the MTM moves at a constant speed, this means individual samples will be spaced $20\,\mu$m apart from one another. In the ideal case, $A\left(\ell\right)$ is even symmetric about the null position, so its spectrum is given by its Fourier cosine transform:

\begin{equation}
A\left(\nu\right)=\int_{-\infty}^{\infty}A\left(\ell\right)\cos\left(2\pi\nu\ell\right)d\ell.
\end{equation}
 
Then the actual spectrum $\tilde{\mathbf{S}}_{x}^{L}\left(\nu\right)$ measured by the detector is given by the convolution of $\mathbf{S}_{x}^{L}\left(\nu\right)$ and $A\left(\nu\right)$:

\begin{equation}\label{eq:spectrum_conv}
\tilde{\mathbf{S}}_{x}^{L}\left(\nu\right)=\mathbf{S}_{x}^{L}\left(\nu\right)\star A\left(\nu\right),
\end{equation}
where the $\star$ operator is the convolution.

The PIXIE mirror scan strategy yields an apodization function given by

\begin{equation}\label{eq:apod_ideal}
A\left(\ell\right)=\left(1-\left(\frac{\ell}{L}\right)^{4}\right)^{2},
\end{equation}
where $L$ is the optical path difference between the two beams at the mirror's maximum displacement from the null. For further details about PIXIE's mirror scanning strategy and the resultant apodization, see \cite{Kogutetal2011}. 

\subsubsection{Asymmetric apodization}\label{sub:apoderr}

The first class of MTM errors we treat comes about if the apodization function $A\left(\ell\right)$ is not symmetric about the null position. In this case, the apodization function becomes

\begin{equation}\label{eq:apodize_err}
\tilde{A}\left(\ell\right)=\left(1-\left(\frac{\ell-\Delta\ell}{L}\right)^{4}\right)^{2},
\end{equation}
where $\Delta\ell$ is is the displacement from the null about which $\tilde{A}\left(\ell\right)$ is symmetric. By the Fourier shift theorem, this gives a corresponding phase shift $\beta\left(\nu\right)$ in the frequency domain, where

\begin{equation}
\beta\left(\nu\right)=-2\pi\nu\Delta\ell.
\end{equation}
Then the spectrum of the apodization function is

\begin{equation}\label{eq:apodiz_err_spec}
\tilde{A}\left(\nu\right)=A\left(\nu\right)\exp\big(i\beta\left(\nu\right)\big).
\end{equation}
 
\begin{figure}[t]
\centering
\includegraphics[width=1\columnwidth]{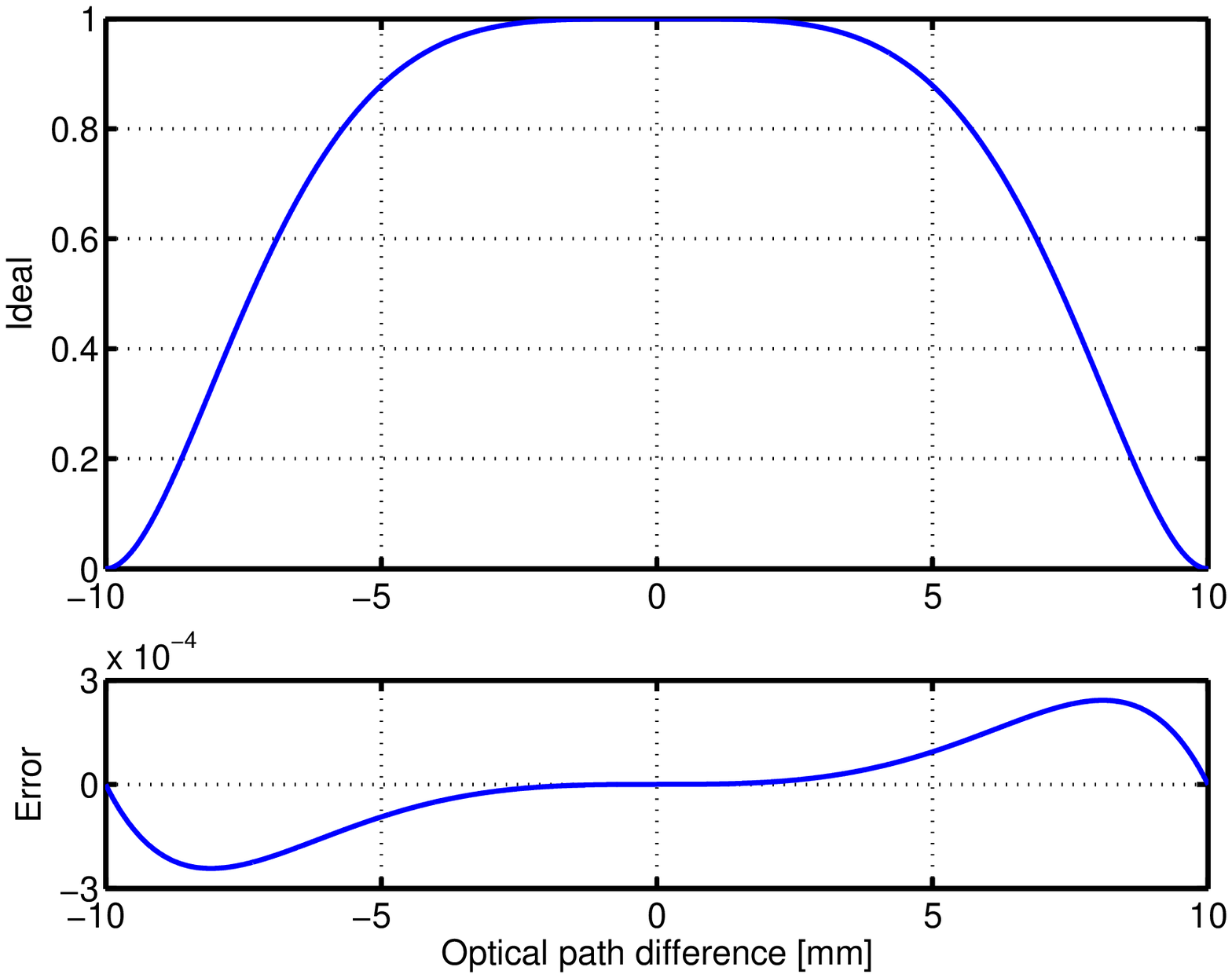}
\includegraphics[width=1\columnwidth]{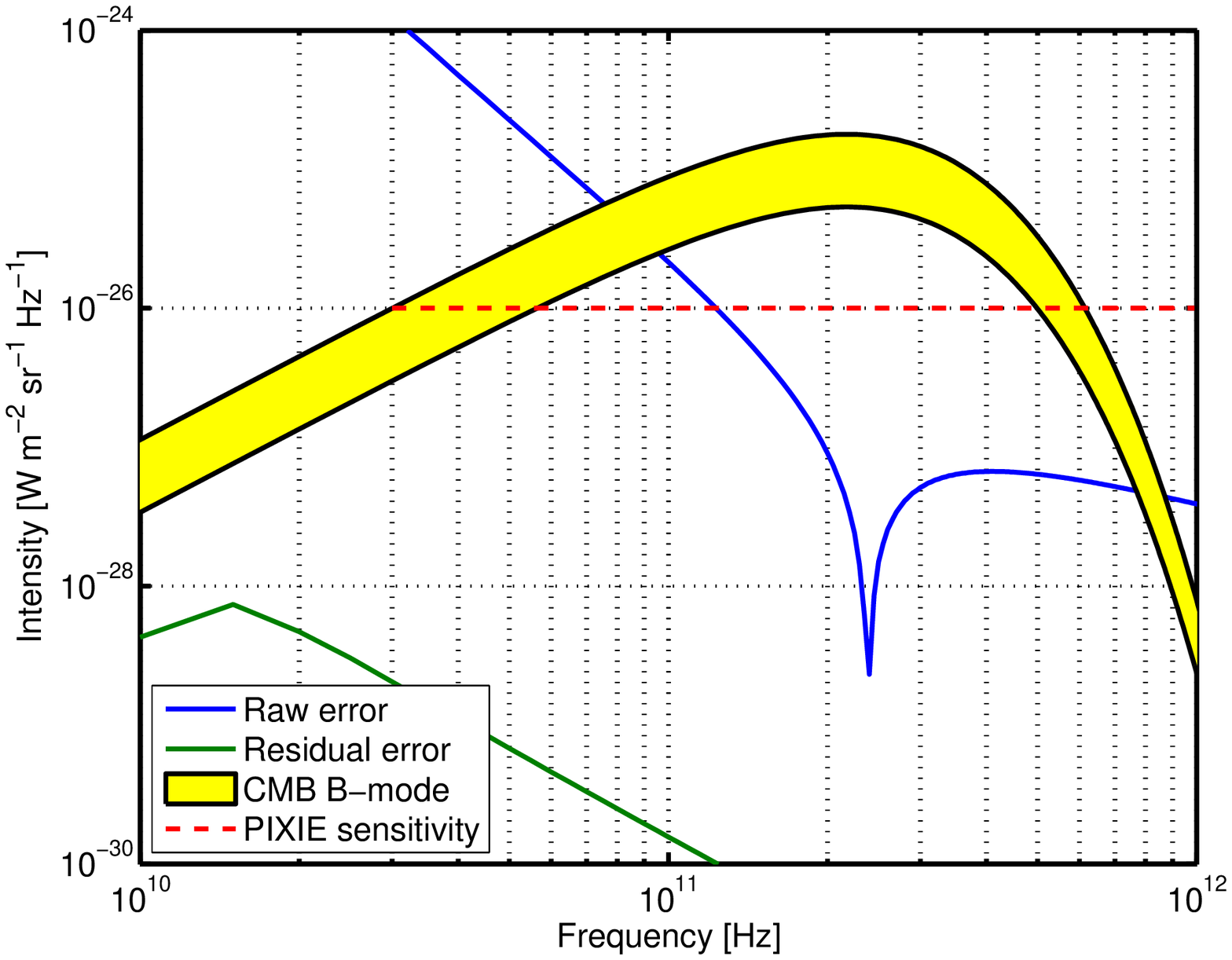}
\caption{Top: Ideal apodization function (top) and its non-ideal counterpart (bottom). Their relative magnitudes are calculated for $\Delta\ell=1$ $\mu$m. Bottom: Error in foreground measurements due to asymmetries in the apodization function. We assume that the mirror stroke is symmetric about a position $\Delta\ell=1$ $\mu$m. Because the error term in the apodization function is nearly odd-symmetric about the mirror's null position, its effects can be mitigated by first averaging the two sides of an interferogram, and then taking the Fourier cosine transform of the averaged one-sided interferogram. The resulting spectrum is many orders of magnitude below the expected {\sl B}-mode signal, which shows amplitudes for $0.01<r<0.1$.}
\label{fig:Apoderr}
\end{figure}

The offset $\Delta\ell$ is small relative to the sample spacing, so we express $\tilde{A}\left(\ell\right)$ as a Taylor series. Keeping terms through second order in $\Delta\ell$, it is
\begin{multline}
\tilde{A}\left(\ell\right)=A\left(\ell\right)+\Delta\ell\left(\frac{8\ell^{3}}{L^{4}}-\frac{8\ell^{7}}{L^{8}}\right)\\-\Delta\ell^{2}\left(\frac{12\ell^{2}}{L^{4}}-\frac{28\ell^{6}}{L^{8}}\right),
\end{multline}
where $A\left(\ell\right)$ is given by Equation \ref{eq:apod_ideal}. This is plotted in Figure \ref{fig:Apoderr}.

The measured spectrum can therefore be expressed as the sum of the ideal apodized spectrum and an error term:

\begin{equation}
\tilde{\mathbf{S}}_{x}^{L}\left(\nu\right)=\mathbf{S}_{x}^{L}\left(\nu\right)\star A\left(\nu\right)+\epsilon_{x}^{L}\left(\nu\right),
\end{equation}
where the error is given by

\begin{multline}
\epsilon_{x}^{L}\left(\nu\right)=\mathbf{S}_{x}^{L}\left(\nu\right)\star \mathcal{F}\Bigg[\Delta\ell\left(\frac{8\ell^{3}}{L^{4}}-\frac{8\ell^{7}}{L^{8}}\right)\\-\Delta\ell^{2}\left(\frac{12\ell^{2}}{L^{4}}-\frac{28\ell^{6}}{L^{8}}\right)\Bigg].
\end{multline}
The operator $\mathcal{F}$ is the Fourier transform.

The effect of the apodization error will be to distort the ideal spectrum. This error can be corrected by averaging the two sides of a full interferogram before taking the Fourier cosine transform of the resultant one sided interferogram. This is equivalent to taking a complex Fourier transform of the full stroke and keeping the real part, which we will do in practice. This works because the error in $\tilde{A}\left(\ell\right)$ is approximately anti-symmetric about the mirror's null position. After performing this operation, apodization error falls to many orders of magnitude below expected {\sl B}-mode signals (see Figure \ref{fig:Apoderr}).

\subsubsection{Asymmetric interferogram}\label{sub:zpd}

If the zeroth sample is not at the null position, then the interferogram will no longer be symmetric about $\ell=0$. In the following, we derive the analytic form for the asymmetric interferogram that results when the zeroth sample is not located at the null. We then introduce the steps that can be taken to symmetrize the interferogram before taking its Fourier transform. We assume that the apodization function $A\left(\ell\right)$ is symmetric about the null and neglect its contribution to the resultant spectrum. For cases where both the interferogram and the apodization function are asymmetric, see \cite{Sakaietal1968}.

If the interferogram does not have a sample at the null position, then we employ the Fourier shift theorem and the power measured by the detector is given by

\begin{equation}\label{eq:interfer_asymm}
\tilde{\mathbf{P}}_{x}^{L}\left(\ell\right)=
\int_{-\infty}^{\infty}\mathbf{S}_{x}^{L}\left(\nu\right)\cos\big(2\pi\nu\ell+\varphi\left(\nu\right)\big)d\nu,
\end{equation}
where $\varphi\left(\nu\right)$ is an asymmetric phase factor that follows $\varphi\left(-\nu\right)=-\varphi\left(\nu\right)$.

For a linear offset from the null position, the phase factor is given by 

\begin{equation}
\varphi\left(\nu\right)=2\pi\nu\Delta\ell,
\end{equation}
where $\Delta\ell$ now corresponds to the displacement of the zeroth sample from the null position. Then the measured power can be re-expressed as

\begin{multline}
\tilde{\mathbf{P}}_{x}^{L}\left(\ell\right)=\int_{-\infty}^{\infty}\mathbf{S}_{x}^{L}\left(\nu\right)\exp\left(-2\pi i\nu\Delta\ell\right)\\\times\exp\left(-2\pi i\nu\ell\right)d\nu.
\end{multline}

The interferogram is no longer symmetric, so to compute the spectrum we must take the complete Fourier transform of the full two-sided interferogram. This gives:

\begin{equation}\label{eq:spec_asymm}
\tilde{\mathbf{S}}_{x}^{L}\left(\nu\right)=\mathbf{S}_{x}^{L}\left(\nu\right)\exp\left(-2\pi i\nu\Delta\ell\right).
\end{equation}

For $\Delta\ell\ll L$, $\tilde{\mathbf{S}}_{x}^{L}\left(\nu\right)$ can be expanded as a Taylor series. Expressed as the sum of the ideal spectrum and associated error terms, it is

\begin{equation}
\tilde{\mathbf{S}}_{x}^{L}\left(\nu\right)=
\mathbf{S}_{x}^{L}\left(\nu\right)+\epsilon_{rx}^{L}\left(\nu\right)+i\epsilon_{ix}^{L}\left(\nu\right),
\end{equation}
where the real and imaginary error terms $\epsilon_{rx}^{L}\left(\nu\right)$ and $\epsilon_{ix}^{L}\left(\nu\right)$, respectively, are given by

\begin{align}
\epsilon_{rx}^{L}\left(\nu\right)&=\mathbf{S}_{x}^{L}\left(\nu\right)\left(2\pi\left(\nu\Delta\ell\right)^{2}\right),\\
\epsilon_{ix}^{L}\left(\nu\right)&=\mathbf{S}_{x}^{L}\left(\nu\right)\left(2\pi\nu\Delta\ell\right).
\end{align}

The real part of the error shows up in second order and is proportional to $\left(\nu\Delta\ell\right)^{2}$. Assuming a null offset $\Delta\ell$ of 10 $\mu$m ({\sl i.e.} $1/2$ sample), this term will be $\sim5$ orders of magnitude smaller than the measured spectrum near the peak brightness of the CMB (see Figure \ref{fig:samperr_null}).

For an ideal FTS, the imaginary component of an interferogram contains no signal, but only instrument noise. In this case, the imaginary part of the error provides a direct measurement of the phase factor $\varphi\left(\nu\right)$. This measurement allows us to symmetrize the measured interferogram such that we can recover the incident spectrum by applying the Fourier cosine transform to half the interferogram. The symmetrization process has important implications with regards to minimizing non-linear noise \citep{Connes1961, Formanetal1966}.

To symmetrize the measured interferogram, we first define $\varphi\left(\ell\right)$ as the Fourier transform of the phase term, where up to some constant,

\begin{equation}
\varphi\left(\ell\right)=\int_{-\infty}^{\infty}\exp\left(2\pi i\nu\Delta\ell\right)\exp\left(-2\pi i\nu\ell\right)d\nu.
\end{equation}

From our measurement of the imaginary component of the error, this will be a known quantity. 

Next we take the convolution of the measured asymmetric interferogram $\tilde{\mathbf{P}}_{x}^{L}\left(\ell\right)$ and the phase term $\varphi\left(\ell\right)$:

\begin{equation}
\bar{\mathbf{P}}_{x}^{L}\left(\ell\right)=\tilde{\mathbf{P}}_{x}^{L}\left(\ell\right)\star\varphi\left(\ell\right).
\end{equation}

$\bar{\mathbf{P}}_{x}^{L}\left(\ell\right)$ is symmetric about the null position, and from its Fourier cosine transform we compute $\mathbf{S}_{x}^{L}\left(\nu\right)$. Assuming that the zeroth sample offset $\Delta\ell$ is, to first order, constant the symmetrization process can be accomplished on a large subset of data from the measurement of a single two-sided interferogram. This process has been demonstrated in practice on measurements made by the FIRAS instrument that flew on the COBE satellite \citep{Fixsenetal1994}. In fact, corrections to FIRAS data were successfully implemented that reduced the effective offset $\Delta\ell$ to the $1$ $\mu$m level. We expect better performance from the lower-noise PIXIE data. 

\subsubsection{Misaligned transfer mirrors}

When the transfer mirrors are misaligned, the phase difference $\phi$ between beams is not well-defined. Instead, the actual phase difference $\tilde{\phi}$ is given by all values in the range $\tilde{\phi}=\phi\pm\delta$, where $2\delta$ is the optical path difference between light hitting opposite extrema of a misaligned optical surface. If the right block of transfer mirrors is misaligned by some small angle $\Delta\theta$ about its vertical axis of symmetry, then the phase factor $\delta$ is given by

\begin{equation}
\delta\simeq r\Delta\theta,
\end{equation}
where $r$ is the radius of the transfer mirror and we used $\sin\Delta\theta\simeq\Delta\theta$.

The measured interferogram $\tilde{\mathbf{P}}_{x}^{L}$ will then be given by the convolution of the ideal interferogram and a square pulse of unity amplitude and width $\delta$. Taking the Fourier transform gives the measured spectrum:

\begin{equation}
\tilde{\mathbf{S}}_{x}^{L}\left(\nu\right)=\sinc\left(2\nu\delta\right)\mathbf{S}_{x}^{L}\left(\nu\right).
\end{equation}

In order to prevent significant attenuation of the sky signal, the 3 dB point of the $\sinc$ function should be at a frequency of $\sim6$ THz. For PIXIE, this corresponds to a misalignment angle $\Delta\theta\simeq35^{\prime\prime}$. Considerably tighter constraints are routinely achieved in FTSs optimized for infrared spectroscopy \citep[e.g.][]{Connes1961}. Likewise for PIXIE-sized mirrors a misalignment angle of $\sim35^{\prime\prime}$ corresponds to machining tolerance of $\sim250$ $\mu$m. This is an order of magnitude larger than typical machine precision.

\subsubsection{Conclusion}

\begin{figure}[t]
\centering
\includegraphics[width=1\columnwidth]{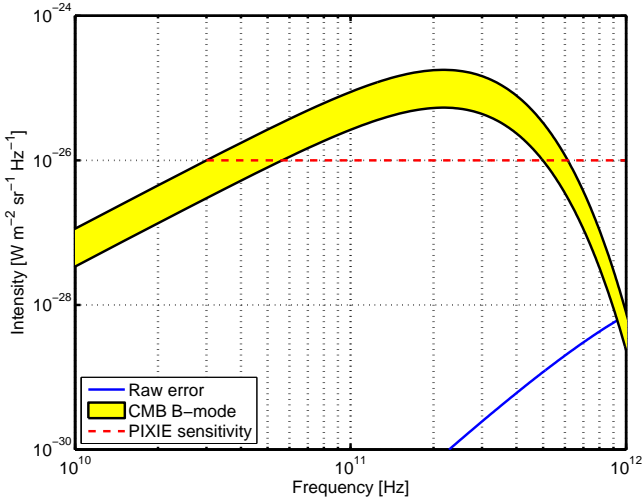}
\caption{Sampling error spectrum due to a zeroth sample offset or instrumental vibrations. While the CMB signal is proportional to $\mathbf{B}^{\prime}_{\nu,T_{0}}$, the error signal is proportional to $\mathbf{B}^{\prime}_{\nu,T_{0}}\nu^{2}$. We assume that the amplitude of the deviations or vibrations is $\Delta\ell=1\,\mu$m. The residual error after corrections is below the scale of this plot. The {\sl B}-mode band shows amplitudes for $0.01<r<0.1$.}
\label{fig:samperr_null}
\end{figure}

We show how phase errors affect the signal measured by PIXIE. Offsets in the apodization function and the position of the zeroth sample lead to assymetries in the interferogram. Misalignments of transfer mirrors attenuate the signal at frequencies defined by the misalignment angle. We correct apodization error by taking the Fourier transform of the full stroke, resulting in a residual error term proportional to some high order in $\Delta\ell$ (Figure \ref{fig:Apoderr}). We correct zeroth sample offsets by measuring the imaginary component of the spectrum and symmetrizing the interferogram prior to taking its Fourier transform. The residual error is orders of magnitude below PIXIE's sensitivity floor (Figure \ref{fig:samperr_null}). The constraint on transfer mirror alignment is weak relative to what is achieved in FTSs optimized for shorter wavelength spectroscopy.

\subsection{Sampling errors}

Sampling errors are systematic or random offsets in the positions of interferogram samples. These can be caused by mechanical uncertainties in the mirror's position or uncertainties in the timing of samples. In the following, we develop a model for sampling errors and quantify their effect on the measured interferogram.

We first determine the maximum position error that allows foreground measurements to better than $0.01\%$ and compare this requirement to the measured performance of a prototype MTM we built. Then following the methods by \cite{Fixsenetal1994}, 
we look at two specific subclasses of sampling errors: errors in PIXIE's absolute frequency scale and uncertainty in the MTM position.

\subsubsection{Position errors}

\begin{figure}[t]
\centering
\includegraphics[width=1\columnwidth]{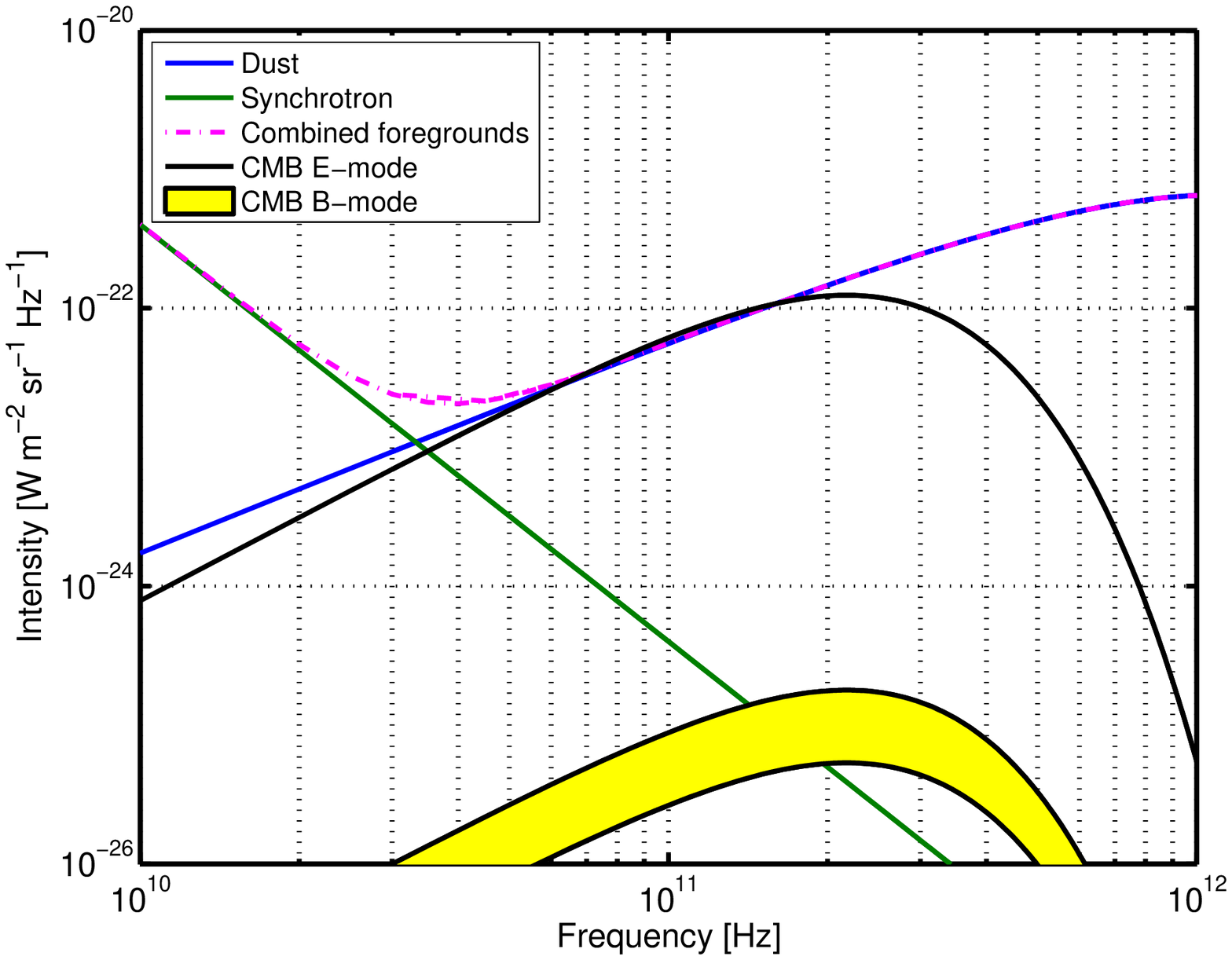}
\includegraphics[width=1\columnwidth]{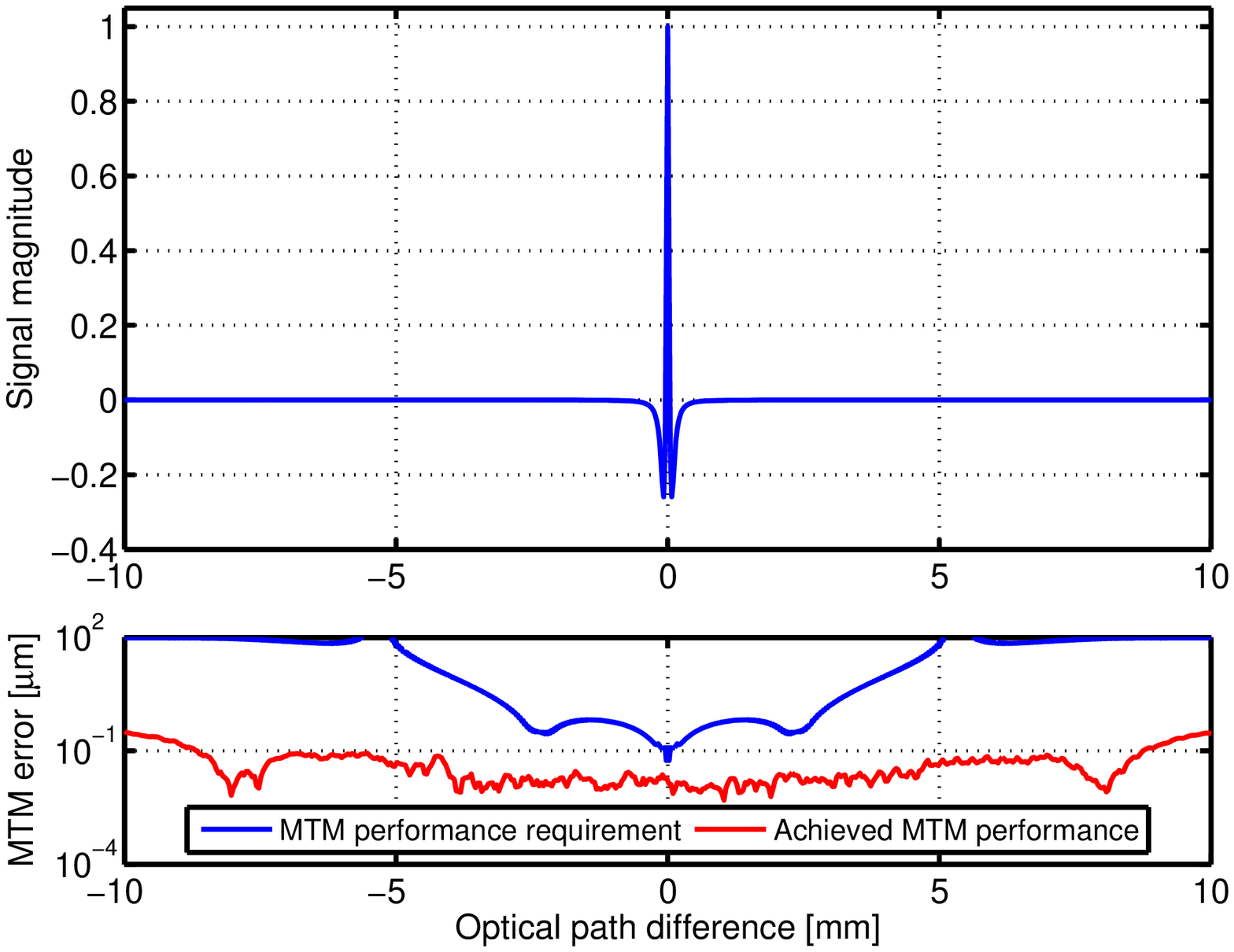}
\caption{Top: Model difference spectrum we expect to measure with PIXIE. The brightness temperature of each component is derived from \cite{Planck2015}. The {\sl B}-mode band shows amplitudes for $0.01<r<0.1$. Bottom: Interferogram showing the Fourier transform of the composite foreground spectrum (top) and the single-pixel MTM position error requirement that gives us $0.01\%$ accuracy in foreground intensity measurements (bottom). The blue curve shows the required performance, and the red curve shows the average measured performance of a prototype MTM that we built.}
\label{fig:PIXIE_modelspec}
\end{figure}

To quantify the magnitude of MTM errors that we can tolerate, we consider a sample difference spectrum that we expect to measure with PIXIE. In the model we include polarized emission from the CMB, galactic synchrotron radiation and thermal dust. Estimates of the brightness temperature of each component are derived from \cite{Planck2015}. The model is shown in Figure \ref{fig:PIXIE_modelspec}. 

\begin{figure}[t]
\centering
\includegraphics[width=1\columnwidth]{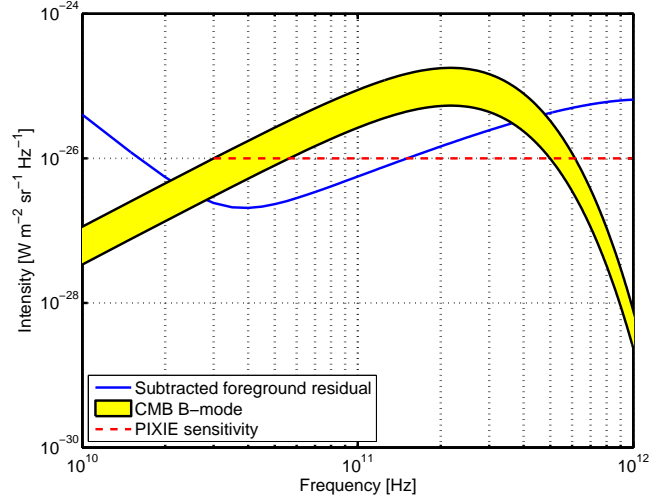}
\caption{Effect of sample position errors on foreground subtractions. We require that sampling errors be small enough that we can measure foregrounds to better than $0.01\%$, enabling accurate foreground subtraction at the peak brightness of the CMB. The {\sl B}-mode band shows amplitudes for $0.01<r<0.1$.}
\label{fig:samperr}
\end{figure}

The Fourier transform of the foreground spectrum gives the interferogram that PIXIE measures. From the derivative of the interferogram, we calculate the maximum MTM position error that enables measurements of the polarized foregrounds to $0.01\%$, enabling accurate subtraction of foregrounds at the frequency of the CMB's peak brightness. This result, along with the average measured performance of the prototype MTM, is shown in Figure \ref{fig:PIXIE_modelspec}. The residual foreground spectrum, after subtracting it to a part in $10^{4}$, is shown in Figure \ref{fig:samperr}.

We do not require sub-$\mu$m position accuracy on every stroke, but rather we require the average error to be small, as in Figure \ref{fig:PIXIE_modelspec}. In addition, because most of the information about the CMB is near the center of the interferogram, we are especially tolerant of position errors near the extrema of the MTM stroke. In our tests, these errors could exceed $2\,\mu$m, but our requirements are an order of magnitude higher.

\subsubsection{Frequency scale errors}

Frequency scale errors can result from thermal contractions of the scale used to determine the MTM's position. They can also result from optical beam divergence in the FTS. Such errors translate to temperature errors in the observed spectrum, and are easily corrected by observing known interstellar emission lines. This technique has been demonstrated in practice; from interstellar line emission calibration, errors in the FIRAS frequency scale were corrected to the $0.1\%$ level \citep{Fixsenetal1994}. This corresponds to position errors of $\sim10^{-3}\,\mu$m, well below our requirements (see Figure \ref{fig:PIXIE_modelspec}).

\subsection{Harmonic oscillations of the instrument}\label{sub:vibs}

In the ideal case, the interferogram measured by the detectors is given by Equation \ref{eq:interogram_ideal}. In the presence of harmonic instrumental vibrations, it becomes

\begin{multline}
\tilde{\mathbf{P}}_{x}^{L}\left(\ell\right)=\int_{-\infty}^{\infty}\mathbf{S}_{x}^{L}\left(\nu\right)\exp\left(-2\pi i\nu\delta\ell\right)\\ \times\exp\left(-2\pi i\nu\ell\right)d\nu.
\end{multline}
Here $\delta\ell=\Delta\ell\sin\left(\omega t+\phi\right)$,
where $\Delta\ell$ is the amplitude of the vibration-induced perturbation to the mirror position, $\omega$ is the angular frequency of the perturbation, and $\phi$ is its phase. The perturbation can either come from vibrations in the MTM's physical position or from sample timing errors since $\ell\left(t\right)=vt$, where $v$ is the speed of the MTM and $t$ is the sample time. Both can result from harmonic vibrations of the instrument \citep{Fixsenetal1994}.

The measured spectrum $\tilde{\mathbf{S}}_{x}^{L}\left(\nu\right)$ is computed by taking the complete Fourier transform of the two-sided interferogram:

\begin{equation}\label{eq:spec_meas_samperr}
\tilde{\mathbf{S}}_{x}^{L}\left(\nu\right)=\mathbf{S}_{x}^{L}\left(\nu\right)\exp\big(-2\pi i\nu\Delta\ell\sin\left(\omega t+\phi\right)\big).
\end{equation}

For small perturbation amplitudes, where $\Delta\ell$ is much smaller than typical sample spacing, the measured spectrum given by Equation \ref{eq:spec_meas_samperr} is expanded as a Taylor series. Keeping terms through second order in $\Delta\ell$, it is

\begin{multline}
\tilde{\mathbf{S}}_{x}^{L}\left(\nu\right)\simeq\mathbf{S}_{x}^{L}\left(\nu\right)\big(1-2\pi i\nu\Delta\ell\sin\left(\omega t+\phi\right)\\-2\pi^{2}\nu^{2}\Delta\ell^{2}\sin^{2}\left(\omega t+\phi\right)\big).
\end{multline}

On average, the imaginary component of the above expression will be zero. Then the measured spectrum is real and can be expressed as the sum of the ideal spectrum and an error term:

\begin{equation}
\mathbf{\tilde{\mathbf{S}}}_{x}^{L}\left(\nu\right)=\mathbf{S}_{x}^{L}\left(\nu\right)+\epsilon_{x}^{L}\left(\nu\right),
\end{equation}
where the error is given by

\begin{equation}
\epsilon_{x}^{L}\left(\nu\right)=-\mathbf{S}_{x}^{L}\left(\nu\right)\left(\pi\nu\Delta\ell\right)^{2}.
\end{equation}

This error shows up in second order in $\Delta\ell$ and will be several orders of magnitude smaller than the ideal signal spectrum. Furthermore, it is proportional to $\nu^{2}$ so it has different spectral content from the sky signal. The solution used to correct this error from FIRAS \citep{Fixsenetal1994} is available for PIXIE data.

\subsubsection{Conclusion}

In this section we consider the role of frequency scale errors and vibration-induced perturbations to the sample location. To correct for the former, we calibrate the frequency scale by observing interstellar line emission. Correcting for the latter to acceptable levels is accomplished by averaging interferograms, causing the imaginary component to drop out and leaving a symmetric interferogram whose spectrum can be computed with the Fourier cosine transform. The results of this analysis are shown in Figure \ref{fig:samperr_null}.

\section{Spin-synchronous errors}\label{sec:Spin-synchronous-effects}

Long term drifts in instruments have many sources, from temperature changes, to radiation damage, to outgassing. Here we are encouraged by the FIRAS experience which included only a single source of drift attributed to the helium defusing out of one of the thermometers on the external calibrator.  Long term drifts on PIXIE can be detected (and corrected) by observing the calibrator. Random drifts are not confused with polarization. Only spin synchronous drifts (specifically those at twice the spin rate) are coupled to the polarization modulation provided by the spin. These could be modulated by thermal effects as the spacecraft rotates in the sunlight, or magnetic effects.

In this section we treat errors which occur at the fundamental frequency or at higher harmonics of the spacecraft's rotation. Polarized light incident on the PIXIE FTS and measured by the detectors will be amplitude modulated at twice the spacecraft's spin frequency (see Appendix \ref{app: Amplitude}). No spin-synchronous errors give rise to amplitude modulated signals that are confused with polarized sky signals. We use the example of spin-synchronous gain drifts to show how spin-synchronous errors show up in the measured signal and are corrected to acceptable levels.

\subsection{Gain drifts}\label{sub:gain}

\begin{figure}[t]
\centering
\includegraphics[width=1\columnwidth]{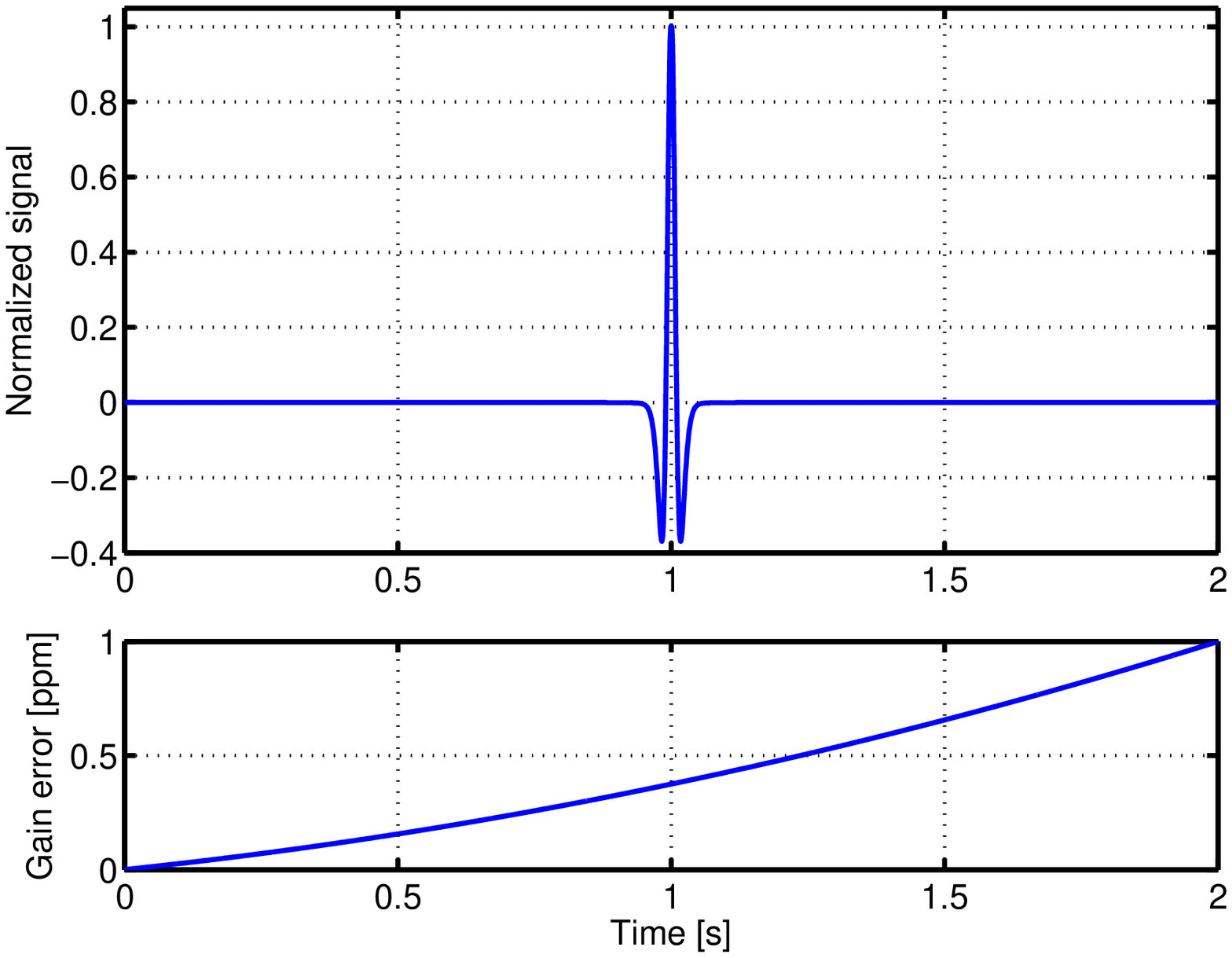}
\includegraphics[width=1\columnwidth]{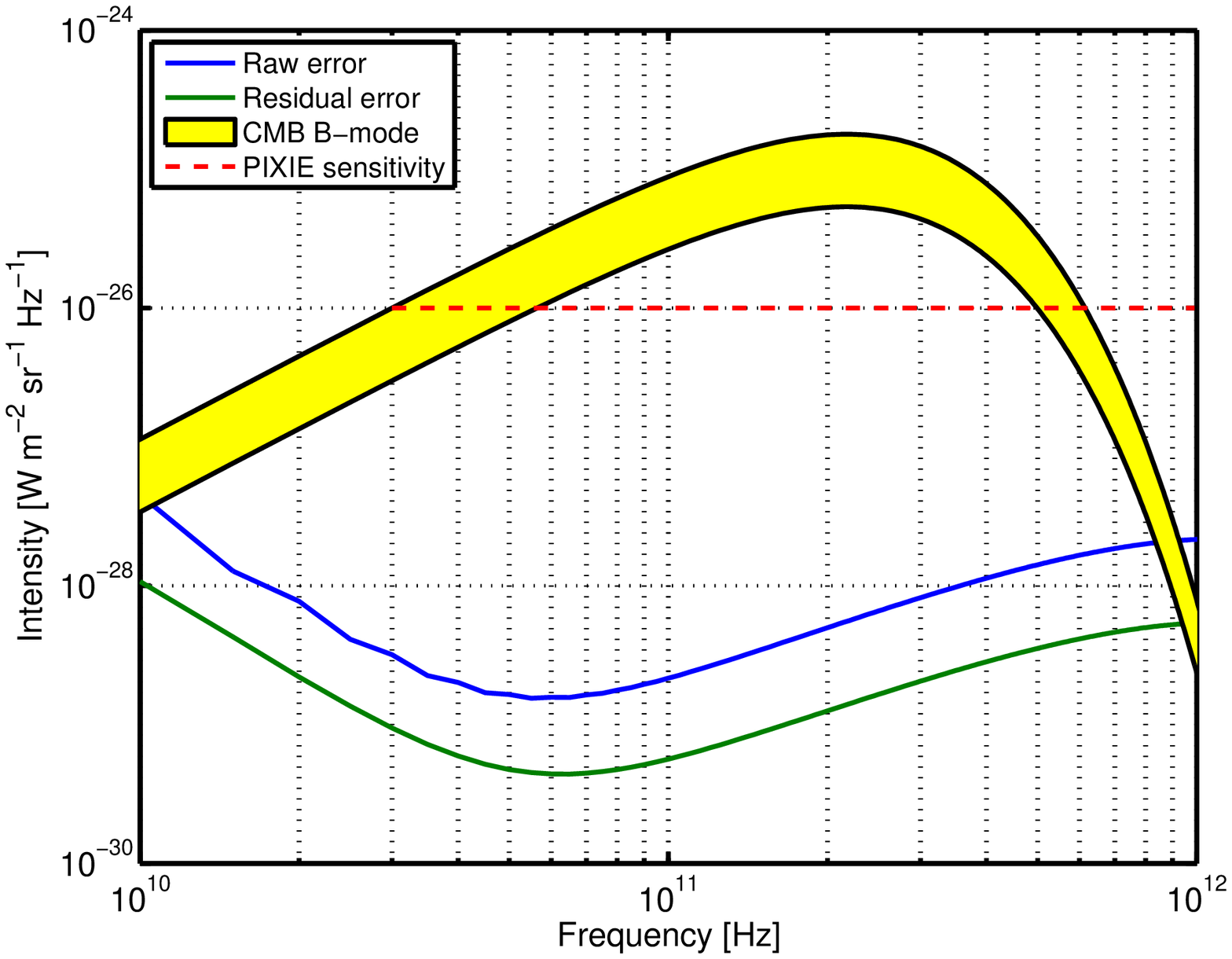}
\caption{Top: Model interferogram exhibiting gain error. This interferogram is the Fourier transform of a typical sky signal multiplied by some gain error. Because most of the information about the CMB is contained in $\sim25\%$ of any given interferogram, a drift $\Delta G/G=1\times10^{-6}$ over the 2 seconds it takes to measure a two-sided interferogram corresponds to a drift $\Delta G/G\simeq1\times10^{-7}$ in each interferogram's critical region. The gain drift is easily corrected by averaging the two sides of the interferogram before taking its Fourier transform, identical to how we correct for asymmetries in the apodization function. Bottom: Error due to spin-synchronous gain drifts. We show the effect of gain errors on foreground subtraction. As expected, the gain errors that multiply the measured interferogram distort the spectrum in the first few frequency bins. The {\sl B}-mode band shows amplitudes for $0.01<r<0.1$.}
\label{fig:Gainerr}
\end{figure}

Spin-synchronous gain drifts can occur, for example, if the temperature of readout electronics varies as the spacecraft spins. As a FTS with multimoded detectors, PIXIE is particularly well equipped to deal with such effects. Imagine a case where the gain $G$ drifts at the instrument's spin frequency. This would occur if the electronics were on one side of the spacecraft, going through one thermal cycle per each $\sim15$ second rotation. Because independent measurements of the sky occur every second, we model such a drift as a low order polynomial that multiplies the measured interferogram. 

An example two-sided interferogram that exhibits a gain drift is shown in Figure \ref{fig:Gainerr}. It shows the Fourier transform of a typical sky signal (Figure \ref{fig:PIXIE_modelspec}) multiplied by a gain drift. We allow a gain drift $\Delta G/G=1\times10^{-6}$ over the two seconds required to take the measurement. This value is based on performance achieved by DMR on COBE. In low Earth orbit, DMR observed no evidence for spin synchronous gain errors at levels greater than $\Delta G/G\simeq1\times10^{-5}$ \citep{Kogutetal1992}. FIRAS published no evidence of spin-synchonous gain drifts, therefore we take DMR's performance to be a relevant published limit. Spinning $\sim4$ times faster than COBE, PIXIE's spin-synchronous gain drifts should be a factor of $\sim4$ smaller for a complete rotation. But one PIXIE measurement takes 2 seconds, so we expect an order of magnitude improvement over DMR for a given measurement.

Before taking corrective actions, we make several observations that illustrate why PIXIE is particularly tolerant of gain errors. First, because periodic gain variations are well approximated by a low order polynomial that multiplies the interferogram, gain error will only show up in the first few frequency bins of the spectrum, away from the peak brightness of the CMB. This is a significant advantage over instruments where different frequency channels with separate detectors and read out schemes are each subject to independent gain drifts. Second, gain errors will affect both polarized and unpolarized signals in the same way, whereas only polarized signals are amplitude modulated by spacecraft rotation. Thus deploying the calibrator enables us to measure gain drifts.

Corrections to gain error are made by averaging two consecutive single-sided interferograms before taking the Fourier transform. This works because the gain curve will average to nearly a constant value over the relevant timescales. For foreground measurements, the residual gain error near the peak brightness of the CMB is at the $10^{-11}$ K level, well below PIXIE's sensitivity floor of a few nK (see Figure \ref{fig:Gainerr}). In practice, we expect to do better than this by measuring the gain curve with the calibrator deployed.

\subsection{Conclusion}

In this section we examined spin-synchronous effects, noting that the absence of amplitude modulation at the frequency of spacecraft rotation means a given signal is not polarized. Taking the specific example of spin-synchronous gain drifts, we show that uncorrected, gain errors give a signal that is much smaller than the expected {\sl B}-mode signal. Averaging a two-sided interferogram before taking the Fourier transform suppresses spin-synchronous gain error by an additional order of magnitude. Other spin-synchronous effects can be treated in a similar way, largely because the rotational period of the instrument ($\sim15$ s) is much longer than the time it takes to make an independent measurement of the CMB ($\sim250$ ms).

\section{Conclusion}

\begin{figure}
\centering
\includegraphics[width=1\columnwidth]{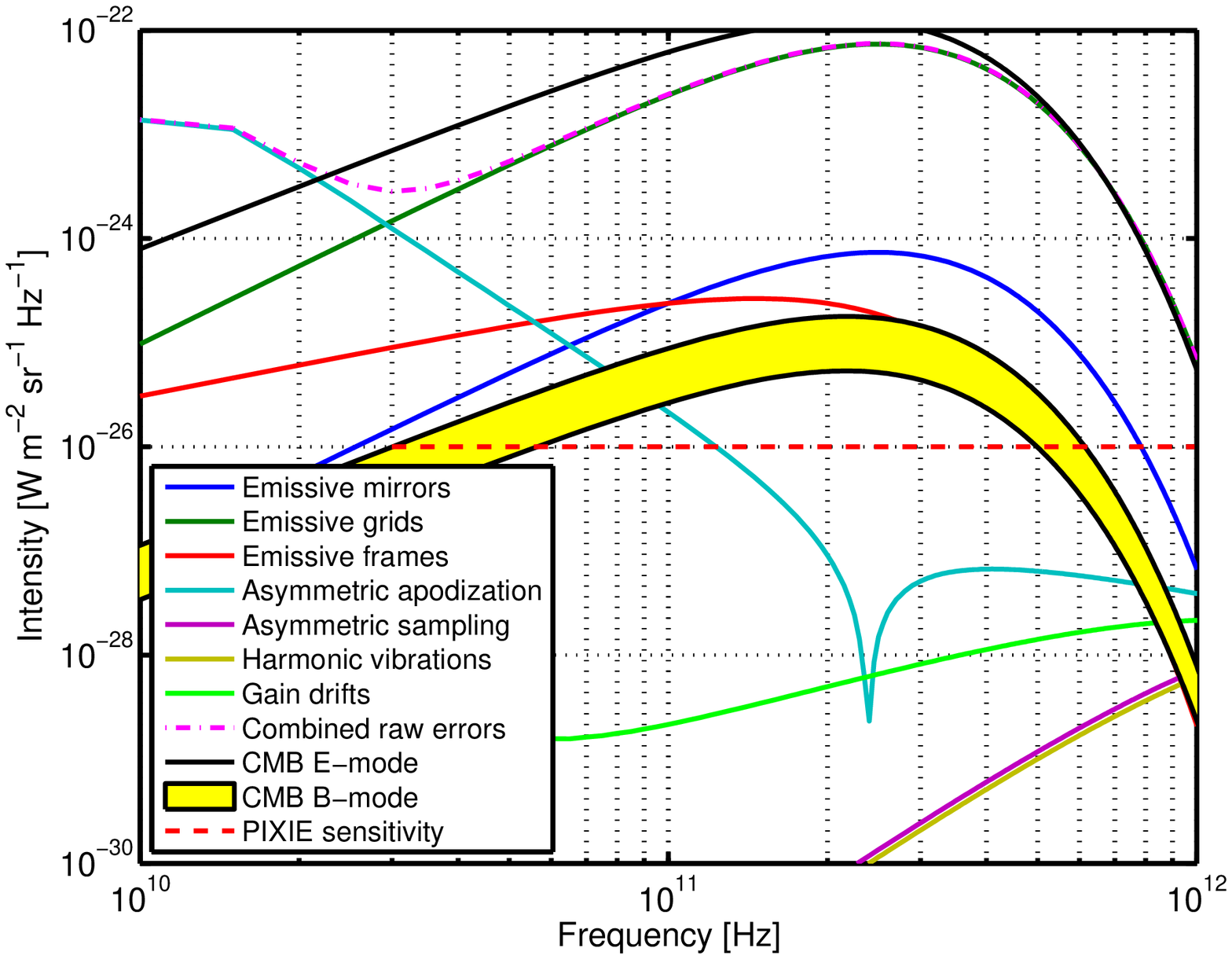}
\includegraphics[width=1\columnwidth]{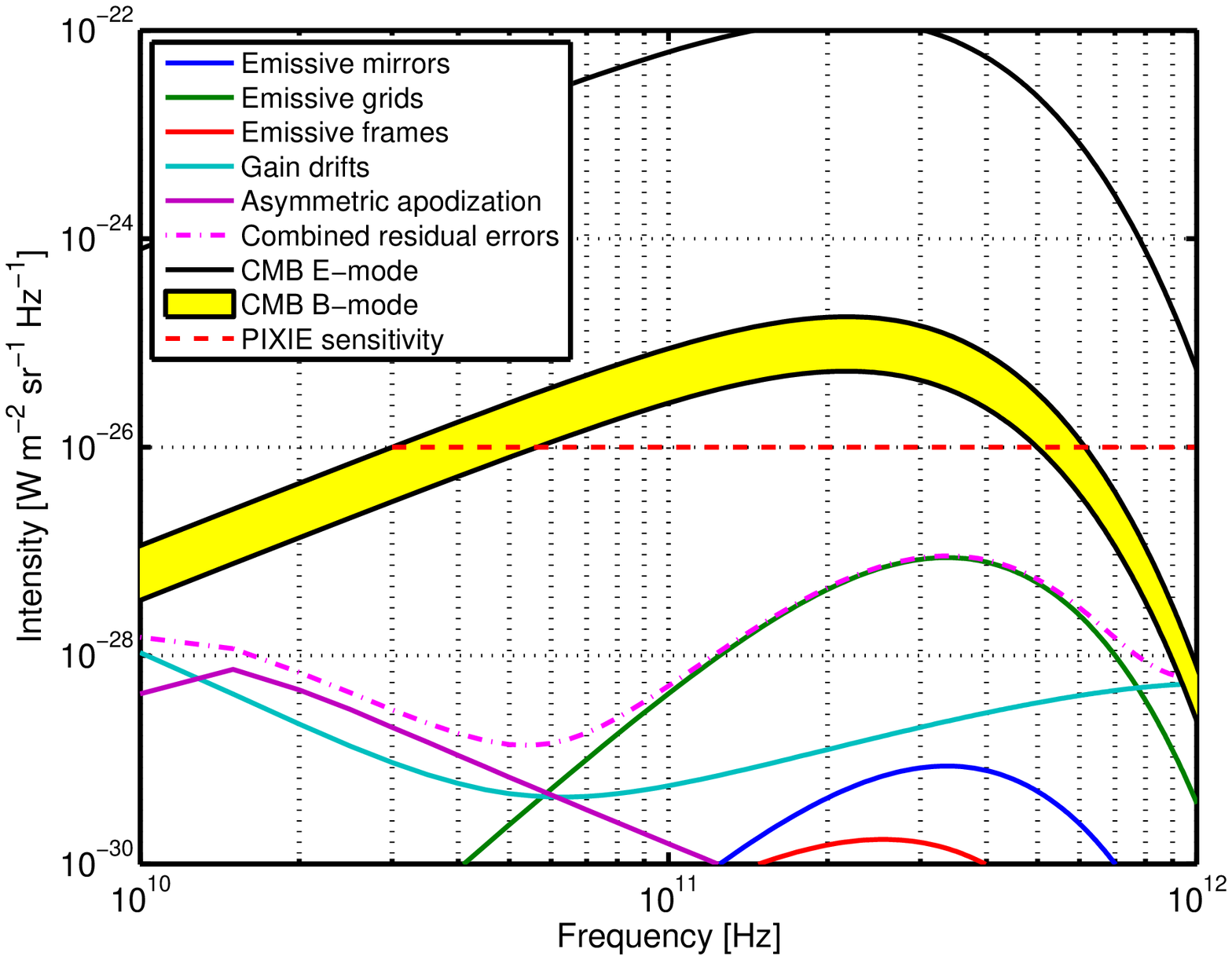}
\caption{Top: Spectra of the raw systematic errors and their combined spectrum. Bottom: Spectra of residual systematic errors and their combined spectrum. Note that the residuals from assymetric sampling and harmonic vibrations are below the scale of this plot. These results are summarized in Table 1. The {\sl B}-mode band shows amplitudes for $0.01<r<0.1$.}
\label{fig: allerr}
\end{figure}

Using PIXIE as an example, we demonstrate how systematic errors in a polarizing FTS designed for CMB observations arise, are identified, and are mitigated. In general, the corrective actions that mitigate systematic errors in PIXIE are simple. Emission errors are generally corrected by actively modulating and controlling the temperature of the optics and taking sums and differences of measured signals. Geometric errors are part of PIXIE's optical efficiency and are measured with the calibrator. Errors from the MTM and spin-synchronous drifts are corrected by taking averages. The errors are easily identified because PIXIE is sensitive to their amplitude and spectral content (see Figure \ref{fig: allerr} and Table 1).

All residual errors are mitigated well below the predicted {\sl B}-mode spectrum without placing unreasonable constraints on PIXIE's design, construction, or observing strategy. PIXIE does not rely on mathematically exact cancellation of potential errors, but instead a series of inherent symmetries that need not be exact when multiplied together. 

An instrument that demonstrates excellent control of systematic effects and has broad frequency coverage is best suited to measure the {\sl B}-mode polarization of the CMB. With a factor of $\sim1000$ fewer detectors and a factor of $\sim100$ more frequency channels than state-of-the-art imagers, an instrument like PIXIE provides a simple, compelling, and proven avenue to accomplish this goal.

\begin{deluxetable}{cccc}[!h]
\tablewidth{0pt}
\tablecolumns{4} 
\tablecaption{Systematic error magnitudes at 215 GHz before and after mitigation}
\tablenum{1}
\tablehead{\colhead{Systematic error} & \colhead{Raw level} & \colhead{Mitigated level} & \colhead{Paper section} \\ 
\colhead{type} & \colhead{(K)} & \colhead{(K)} & \colhead{} } 

\startdata
Emissive mirrors & $10^{-7}$ & $10^{-12}$ & \ref{sub:Primary_nze}  \\

Emissive grids & $10^{-5}$ & $10^{-10}$ & \ref{sub:Non-ideal-polarizing-grid} \\

Emissive frames & $10^{-7}$ & $10^{-12}$ & \ref{sub:no} \\

Assymetric apodization & $10^{-10}$ & $10^{-13}$ & \ref{sub:apoderr} \\

Assymetric sampling & $10^{-13}$ & $10^{-15}$ & \ref{sub:zpd} \\

Harmonic vibrations & $10^{-13}$ & $10^{-15}$ & \ref{sub:vibs} \\

Gain drifts & $10^{-11}$ & $10^{-12}$ & \ref{sub:gain}
\enddata
\end{deluxetable}

\bibliographystyle{apj}
\bibliography{nagler-2015-syserr}

\begin{thebibliography}{}
\expandafter\ifx\csname natexlab\endcsname\relax\def\natexlab#1{#1}\fi

\bibitem[{{Andr{\'e}} {et~al.}(2014){Andr{\'e}}, {Baccigalupi}, {Banday},
  {Barbosa}, {Barreiro}, {Bartlett}, {Bartolo}, {Battistelli}, {Battye},
  {Bendo}, {Beno{\^i}t}, {Bernard}, {Bersanelli}, {B{\'e}thermin}, {Bielewicz},
  {Bonaldi}, {Bouchet}, {Boulanger}, {Brand}, {Bucher}, {Burigana}, {Cai},
  {Camus}, {Casas}, {Casasola}, {Castex}, {Challinor}, {Chluba}, {Chon},
  {Colafrancesco}, {Comis}, {Cuttaia}, {D'Alessandro}, {Da Silva}, {Davis}, {de
  Avillez}, {de Bernardis}, {de Petris}, {de Rosa}, {de Zotti}, {Delabrouille},
  {D{\'e}sert}, {Dickinson}, {Diego}, {Dunkley}, {En{\ss}lin}, {Errard},
  {Falgarone}, {Ferreira}, {Ferri{\`e}re}, {Finelli}, {Fletcher}, {Fosalba},
  {Fuller}, {Galli}, {Ganga}, {Garc{\'{\i}}a-Bellido}, {Ghribi}, {Giard},
  {Giraud-H{\'e}raud}, {Gonzalez-Nuevo}, {Grainge}, {Gruppuso}, {Hall},
  {Hamilton}, {Haverkorn}, {Hernandez-Monteagudo}, {Herranz}, {Jackson},
  {Jaffe}, {Khatri}, {Kunz}, {Lamagna}, {Lattanzi}, {Leahy}, {Lesgourgues},
  {Liguori}, {Liuzzo}, {Lopez-Caniego}, {Macias-Perez}, {Maffei}, {Maino},
  {Mangilli}, {Martinez-Gonzalez}, {Martins}, {Masi}, {Massardi}, {Matarrese},
  {Melchiorri}, {Melin}, {Mennella}, {Mignano}, {Miville-Desch{\^e}nes},
  {Monfardini}, {Murphy}, {Naselsky}, {Nati}, {Natoli}, {Negrello}, {Noviello},
  {O'Sullivan}, {Paci}, {Pagano}, {Paladino}, {Palanque-Delabrouille},
  {Paoletti}, {Peiris}, {Perrotta}, {Piacentini}, {Piat}, {Piccirillo},
  {Pisano}, {Polenta}, {Pollo}, {Ponthieu}, {Remazeilles}, {Ricciardi},
  {Roman}, {Rosset}, {Rubino-Martin}, {Salatino}, {Schillaci}, {Shellard},
  {Silk}, {Starobinsky}, {Stompor}, {Sunyaev}, {Tartari}, {Terenzi},
  {Toffolatti}, {Tomasi}, {Trappe}, {Tristram}, {Trombetti}, {Tucci}, {Van de
  Weijgaert}, {Van Tent}, {Verde}, {Vielva}, {Wandelt}, {Watson}, \&
  {Withington}}]{Andreetal2014}
{Andr{\'e}}, P., {Baccigalupi}, C., {Banday}, A., {et~al.} 2014, \jcap, 2, 6

\bibitem[{{Aumont} {et~al.}(2010){Aumont}, {Conversi}, {Thum}, {Wiesemeyer},
  {Falgarone}, {Mac{\'{\i}}as-P{\'e}rez}, {Piacentini}, {Pointecouteau},
  {Ponthieu}, {Puget}, {Rosset}, {Tauber}, \& {Tristram}}]{Aumontetal2010}
{Aumont}, J., {Conversi}, L., {Thum}, C., {et~al.} 2010, \aap, 514, A70

\bibitem[{{Bell}(1972)}]{Bell1972}
{Bell}, R.~J. 1972, {Introductory Fourier transform spectroscopy}

\bibitem[{{BICEP2/Keck and Planck Collaborations} {et~al.}(2015){BICEP2/Keck
  and Planck Collaborations}, {Ade}, {Aghanim}, {Ahmed}, {Aikin}, {Alexander},
  {Arnaud}, {Aumont}, {Baccigalupi}, {Banday}, \& et~al.}]{BICEP2015}
{BICEP2/Keck and Planck Collaborations}, {Ade}, P.~A.~R., {Aghanim}, N.,
  {et~al.} 2015, Physical Review Letters, 114, 101301

\bibitem[{{Bock} {et~al.}(1995){Bock}, {Lange}, {Parikh}, \&
  {Fischer}}]{Bocketal1995}
{Bock}, J.~J., {Lange}, A.~E., {Parikh}, M.~K., \& {Fischer}, M.~L. 1995, \ao,
  34, 4812

\bibitem[{{Born} \& {Wolf}(1999)}]{BornandWolf1999}
{Born}, M., \& {Wolf}, E. 1999, {Principles of Optics}

\bibitem[{{Chuss} {et~al.}(2012){Chuss}, {Wollack}, {Henry}, {Hui}, {Juarez},
  {Krejny}, {Moseley}, \& {Novak}}]{Chussetal2012}
{Chuss}, D.~T., {Wollack}, E.~J., {Henry}, R., {et~al.} 2012, Applied Optics,
  51, 197

\bibitem[{Connes(1961)}]{Connes1961}
Connes, J.~R. 1961, Recherches sur la spectroscopie par transformations de
  Fourier ({\'E}d. de la" Revue d'optique th{\'e}orique et instrumentale)

\bibitem[{{Fixsen} {et~al.}(1996){Fixsen}, {Cheng}, {Gales}, {Mather},
  {Shafer}, \& {Wright}}]{Fixsenetal1996}
{Fixsen}, D.~J., {Cheng}, E.~S., {Gales}, J.~M., {et~al.} 1996, \apj, 473, 576

\bibitem[{{Fixsen} {et~al.}(1994){Fixsen}, {Cheng}, {Cottingham}, {Eplee},
  {Hewagama}, {Isaacman}, {Jensen}, {Mather}, {Massa}, {Meyer}, {Noerdlinger},
  {Read}, {Rosen}, {Shafer}, {Trenholme}, {Weiss}, {Bennett}, {Boggess},
  {Wilkinson}, \& {Wright}}]{Fixsenetal1994}
{Fixsen}, D.~J., {Cheng}, E.~S., {Cottingham}, D.~A., {et~al.} 1994, \apj, 420,
  457

\bibitem[{{Forman} {et~al.}(1966){Forman}, {Steel}, \&
  {Vanasse}}]{Formanetal1966}
{Forman}, M.~L., {Steel}, W.~H., \& {Vanasse}, G.~A. 1966, Journal of the
  Optical Society of America (1917-1983), 56, 59

\bibitem[{{Gush} {et~al.}(1990){Gush}, {Halpern}, \& {Wishnow}}]{Gushetal1990}
{Gush}, H.~P., {Halpern}, M., \& {Wishnow}, E.~H. 1990, Physical Review
  Letters, 65, 537

\bibitem[{{Hu} {et~al.}(2003){Hu}, {Hedman}, \& {Zaldarriaga}}]{Huetal2003}
{Hu}, W., {Hedman}, M.~M., \& {Zaldarriaga}, M. 2003, \prd, 67, 043004

\bibitem[{{Jones}(1941)}]{Jones1941}
{Jones}, R.~C. 1941, Journal of the Optical Society of America (1917-1983), 31,
  488

\bibitem[{{Kamionkowski} {et~al.}(1997){Kamionkowski}, {Kosowsky}, \&
  {Stebbins}}]{Kamionkowskietal1997}
{Kamionkowski}, M., {Kosowsky}, A., \& {Stebbins}, A. 1997, Physical Review
  Letters, 78, 2058

\bibitem[{{Kogut} {et~al.}(1992){Kogut}, {Smoot}, {Bennett}, {Wright}, {Aymon},
  {de Amici}, {Hinshaw}, {Jackson}, {Kaita}, {Keegstra}, {Lineweaver},
  {Loewenstein}, {Rokke}, {Tenorio}, {Boggess}, {Cheng}, {Gulkis}, {Hauser},
  {Janssen}, {Kelsall}, {Mather}, {Meyer}, {Moseley}, {Murdock}, {Shafer},
  {Silverberg}, {Weiss}, \& {Wilkinson}}]{Kogutetal1992}
{Kogut}, A., {Smoot}, G.~F., {Bennett}, C.~L., {et~al.} 1992, \apj, 401, 1

\bibitem[{{Kogut} {et~al.}(2011){Kogut}, {Fixsen}, {Chuss}, {Dotson}, {Dwek},
  {Halpern}, {Hinshaw}, {Meyer}, {Moseley}, {Seiffert}, {Spergel}, \&
  {Wollack}}]{Kogutetal2011}
{Kogut}, A., {Fixsen}, D.~J., {Chuss}, D.~T., {et~al.} 2011, \jcap, 7, 25

\bibitem[{{Kogut} {et~al.}(2014){Kogut}, {Chuss}, {Dotson}, {Dwek}, {Fixsen},
  {Halpern}, {Hinshaw}, {Meyer}, {Moseley}, {Seiffert}, {Spergel}, \&
  {Wollack}}]{Kogutetal2014}
{Kogut}, A., {Chuss}, D.~T., {Dotson}, J., {et~al.} 2014, in Society of
  Photo-Optical Instrumentation Engineers (SPIE) Conference Series, Vol. 9143,
  Society of Photo-Optical Instrumentation Engineers (SPIE) Conference Series,
  1

\bibitem[{{Mertz}(1967)}]{Mertz1967}
{Mertz}, L. 1967, Infrared Physics, 7, 17

\bibitem[{{Mueller}(1943)}]{Mueller1943}
{Mueller}, H. 1943, {Memorandum on the polarization optics of the photo-elastic
  shutter}, Tech. rep.

\bibitem[{{Planck Collaboration} {et~al.}(2015){Planck Collaboration}, {Adam},
  {Ade}, {Aghanim}, {Alves}, {Arnaud}, {Ashdown}, {Aumont}, {Baccigalupi},
  {Banday}, \& et~al.}]{Planck2015}
{Planck Collaboration}, {Adam}, R., {Ade}, P.~A.~R., {et~al.} 2015, ArXiv
  e-prints, arXiv:1502.01588

\bibitem[{{Sakai} {et~al.}(1968){Sakai}, {Vanasse}, \&
  {Forman}}]{Sakaietal1968}
{Sakai}, H., {Vanasse}, G.~A., \& {Forman}, M.~L. 1968, Journal of the Optical
  Society of America (1917-1983), 58, 84

\bibitem[{{Seljak} \& {Zaldarriaga}(1997)}]{Seljaketal1997}
{Seljak}, U., \& {Zaldarriaga}, M. 1997, Physical Review Letters, 78, 2054

\bibitem[{{Spencer} {et~al.}(2011){Spencer}, {Naylor}, {Ade}, \&
  {Zhang}}]{Spenceretal2011}
{Spencer}, L.~D., {Naylor}, D.~A., {Ade}, P.~A.~R., \& {Zhang}, J. 2011,
  Journal of the Optical Society of America A, 28, 1805

\bibitem[{{Tucker} {et~al.}(1997){Tucker}, {Gush}, {Halpern}, {Shinkoda}, \&
  {Towlson}}]{Tuckeretal1997}
{Tucker}, G.~S., {Gush}, H.~P., {Halpern}, M., {Shinkoda}, I., \& {Towlson}, W.
  1997, \apjl, 475, L73

\bibitem[{Wolf(1959)}]{Wolf1959}
Wolf, E. 1959, Il Nuovo Cimento, 13, 1165

\bibitem[{{Woody} \& {Richards}(1981)}]{Woodyetal1981}
{Woody}, D.~P., \& {Richards}, P.~L. 1981, \apj, 248, 18

\end{thebibliography}

\appendix

\section{Amplitude modulation of polarized light}\label{app: Amplitude}

Here we review how linearly polarized signals incident on the PIXIE FTS become amplitude modulated when we include spacecraft rotation. Complete derivations are given by \cite{Kogutetal2011, Kogutetal2014}, so some details are omitted here. 

In Section \ref{sub:Ideal}, we computed the power measured by the ideal instrument in instrument-fixed coordinates. With both sides of the instrument open to the sky, and for the left side $\hat{x}$ detector, this is given by

\begin{equation}\label{eq:Plx_fringes}
\mathbf{P}_{x}^{L}=\frac{1}{2}\int\left(E_{x}^{sky\,2}-E_{y}^{sky\,2}\right)\cos\left(\frac{4\nu z}{c}\right)d\nu,
\end{equation}
where light incident on the instrument is represented by the vector $E^{sky}=E^{sky}_{x}\hat{x}+E^{sky}_{y}\hat{y}$.

When we include spacecraft rotation, it is necessary to transform from instrument-fixed to sky-fixed coordinates. These transformations take the following form:

\begin{align}
E_{x}^{sky}\hat{x}&\rightarrow\left(E_{x}^{sky}\cos\gamma+E_{y}^{sky}\sin\gamma\right)\hat{x}^{\prime},\\
E_{y}^{sky}\hat{y}&\rightarrow\left(E_{y}^{sky}\cos\gamma-E_{x}^{sky}\sin\gamma\right)\hat{y}^{\prime},
\end{align}
where $\left(\hat{x},\hat{y}\right)$ are instrument-fixed unit vectors, $\left(\hat{x}^{\prime},\hat{y}^{\prime}\right)$ are sky-fixed unit vectors, and $\gamma$ is the spacecraft rotation angle. 

With this transformation, Equation \ref{eq:Plx_fringes} becomes

\begin{equation}\label{Plx_fringes-QU}
\mathbf{P}_{x}^{L}=\frac{1}{2}\int\left(Q^{sky}\cos\left(2\gamma\right)+U^{sky}\sin\left(2\gamma\right)\right)\cos\left(\frac{4\nu z}{c}\right)d\nu,
\end{equation}
where $Q^{sky}$ and $U^{sky}$ are the Stokes $Q$ and $U$ parameters of the sky, given by $\left\langle E_{x}^{sky\,2}-E_{y}^{sky\,2}\right\rangle$ and $2\re\left\langle E_{x}^{sky}E_{y}^{sky}\right\rangle$, respectively. Linearly polarized sky light described by the Stokes parameters $Q$ and $U$ is amplitude modulated at twice the spacecraft's spin frequency. Any signal that is not amplitude modulated at twice the rotation frequency is not from a polarized sky source.

\section{PIXIE as a two-element radio array}

The standard treatment of PIXIE's measured signal (Equation \ref{eq:Plx_ideal}) ignores the spatial separation between its two primary mirrors. Taking the separation into account, we may think of PIXIE as a two-element radio interferometer. Here we model PIXIE as such, solving for its interferometric response to the Stokes parameters of the sky. We show that there is no interferometric reponse to Stokes $I$ and $Q$ in instrument-fixed coordinates. The interferometric response to Stokes $U$ and $V$ in instrument-fixed coordinates shows up in the Fourier sine transform of the measured interferogram. The response, however, is highly oscillatory and is anti-symmetric with regard to the beam pattern, so large scale cancellation will occur when integrating over PIXIE's tophat beam pattern. In addition, because the interferometric response is limited to the imaginary component of the spectrum, it will not interfere with estimates of the ideal signal.

Light from a common source will travel a different path length to the two mirrors. This path difference gives rise to a phase difference between light incident on the mirrors. If we define the baseline separation between the mirrors as the vector $\mathbf{b}$, and the source position on the sky as the vector $\mathbf{s}$, then the phase factor $\beta\left(\nu\right)$ between beams is given by

\begin{equation}
\beta\left(\nu\right)=\nu\tau,
\end{equation}
where $\tau$ is given by
\begin{equation}
\tau=\mathbf{b}\cdot\mathbf{s}/c.
\end{equation}

With both beams open to the sky, the vectors $E^{L}$ and $E^{R}$ describe light incident on the left and right sides of the instrument, respectively:

\begin{equation}
\begin{aligned}
E^{L}&=\mathscr{A}\hat{x}+\mathscr{B}\hat{y},\\ E^{R}&=\mathscr{A}\exp\left(i\nu\tau\right)\hat{x}+\mathscr{B}\exp\left(i\nu\tau\right)\hat{y}.
\end{aligned}
\end{equation}

The power measured by the left side $\hat{x}$ detector is then:

\begin{equation}\label{eq:Plx_tli}
\tilde{\mathbf{P}}_{x}^{L}=\frac{1}{2}\int\Bigg(\left(\mathscr{A}^{2}+\mathscr{B}^{2}\right)+\left(\mathscr{A}^{2}-\mathscr{B}^{2}\right)\cos\left(\frac{4\nu z}{c}\right)+\big(\im\left(\mathscr{A}\mathscr{B}^{\star}\right)\cos\left(\nu\tau\right)-\re\left(\mathscr{A}\mathscr{B}^{\star}\big)\sin\left(\nu\tau\right)\right)\sin\left(\frac{4\nu z}{c}\right)\Bigg)d\nu.
\end{equation}

Expressing Equation \ref{eq:Plx_tli} in terms of the Stokes parameters, we get:

\begin{equation}\label{eq:Plx_tli_stokes}
\tilde{\mathbf{P}}_{x}^{L}=\frac{1}{2}\int\Bigg(I+Q\cos\left(\frac{4\nu z}{c}\right)+\big(V\cos\left(\nu\tau\right)-U\sin\left(\nu\tau\right)\big)\sin\left(\frac{4\nu z}{c}\right)\Bigg)d\nu,
\end{equation}
where the Stokes parameters $I$, $Q$, $U$, and $V$ are given by:
\begin{equation}
\begin{aligned}
I&=\left\langle\mathscr{A}^{2}+\mathscr{B}^{2}\right\rangle,\\
Q&=\left\langle\mathscr{A}^{2}-\mathscr{B}^{2}\right\rangle,\\
U&=2\re\left\langle\mathscr{A}\mathscr{B}^{\star}\right\rangle,\\
V&=2\im\left\langle\mathscr{A}\mathscr{B}^{\star}\right\rangle.
\end{aligned}
\end{equation}

Equation \ref{eq:Plx_tli_stokes} consists of a DC term, a term modulated by the cosine of the mirror movement (the ideal interferogram), and a term modulated by the sine of the mirror movement (the interferometric response). Several factors combine to make the interferometric response term small. The millimeter sky is dominated by the CMB and diffuse dust cirrus, neither of which is thought to emit circular polarization, so we can neglect the term proportional to Stokes $V$. The term proportional to Stokes $U$ will be dominated by CMB {\sl E}-mode polarization and diffuse foreground emission. To estimate its magnitude, we integrate the Stokes $U$ term over the primary beam pattern.

Since the baseline $\mathbf{b}$ separating the two mirrors is normal to the antenna boresight, the phase lag $\tau$ will vanish for a point source on axis and it is anti-symmetric for off-axis sources. Expanding the sky signal in a polar coordinate system $\left[\theta,\phi\right]$ centered on the antenna boresight, the phase term is re-expressed as:

\begin{equation}\label{eq:nutau_polar}
\begin{aligned}
\nu\tau&=2\pi\frac{b}{\lambda}\theta\cos\left(\phi\right)\\&=\frac{\nu b}{c}\theta\cos\left(\phi\right),
\end{aligned}
\end{equation}
where $\theta$ is the radial distance from the boresight, $\phi$ is the angle from the centerline connecting the two primary mirrors, and $\lambda$ is the wavelength of light.

Substituting the polar coordinate phase term (Equation \ref{eq:nutau_polar}) into Equation \ref{eq:Plx_tli_stokes} and integrating over the beam, we get the following response to Stokes $U$:

\begin{equation}\label{eq:Plx_U}
\tilde{\mathbf{P}}_{x}^{L}\big|_{U}=\frac{1}{2}\int\Bigg(\int_{beam}U\left(\nu,\theta,\phi\right)\sin\left(\frac{2\pi\nu b}{c}\theta\cos\left(\phi\right)\right)d\theta d\phi\Bigg)\sin\left(\frac{4\nu z}{c}\right)d\nu.
\end{equation}

The beam integral in Equation \ref{eq:Plx_U} is frequency dependent, with the ratio $b/\lambda$ ranging from $b/\lambda=60$ at $\lambda=1$ cm to $b/\lambda=12000$ at $\lambda=50$ $\mu$m. Then the phase term is highly oscillatory across the $\theta=1.1^{\circ}$ tophat radius, so the sky structure $U\left(\theta,\phi\right)$ will largely cancel in the beam integral. In addition, we can totally eliminate PIXIE's sensitivity to either Stokes $U$ or $V$ by deploying the calibrator.

\end{document}